\documentclass[aps,twocolumn,amsmath,amssymb,preprintnumbers]{revtex4}
\usepackage{amsmath} \usepackage{amsfonts} \usepackage{amssymb}
\usepackage{bbm}
\usepackage{epsfig}
\usepackage{graphics}
\usepackage{graphicx}
\newcommand{\be}{\begin{equation}}
\newcommand{\ee}{\end{equation}}
\newcommand{\ba}{\begin{eqnarray}}
\newcommand{\ea}{\end{eqnarray}}
\newcommand{\nn}{\nonumber}
\newcommand{\kr}{\rangle}
\newcommand{\kl}{\langle}
\newcommand{\m}{(x;\mu)}
\newcommand{\tr}{{\rm tr}}
\newcommand{\s}{{\cal S}}
\newcommand{\rchi}{{\mathpalette\irchi\relax}}
\newcommand{\irchi}[2]{\raisebox{\depth}{$#1\chi$}} % inner command, used by \rchi

\begin{document}

\title[ ]{Scalar lattice gauge theory}

\author{C. Wetterich}
\affiliation{Institut  f\"ur Theoretische Physik\\
Universit\"at Heidelberg\\
Philosophenweg 16, D-69120 Heidelberg}

\begin{abstract}
Scalar lattice gauge theories are models for scalar fields with local gauge symmetries. No fundamental gauge fields, or link variables in a lattice regularization, are introduced.
The latter rather emerge as collective excitations composed from scalars. 
For suitable parameters scalar lattice gauge theories lead to confinement, with all continuum observables identical to usual lattice gauge theories. 
These models or their fermionic counterpart may be helpful for a realization of gauge theories by ultracold atoms.
We conclude that the gauge bosons of the standard model of particle physics  can arise as collective fields within models formulated for other ``fundamental'' degrees of freedom. 
\end{abstract}

\maketitle

\section{Introduction}
\label{Introduction}
Gauge bosons are a central building block of the standard model of particle physics. 
Are those gauge bosons necessarily fundamental, or could they be composites of other fundamental degrees of freedom? 
In the high momentum regime of asymptotically free non-abelian gauge theories the gauge bosons are effectively massless particles. 
If an underlying theory with different degrees of freedom, as scalars or fermions, is to produce massless collective states there needs to be a reason for this. 
Indeed, collective massless spin-one bosons may be due to a local gauge symmetry. 
In the perturbative regime this forbids a mass term for the gauge bosons. 
We therefore investigate models with a local gauge symmetry, but without introducing gauge bosons as fundamental degrees of freedom. 

Composite gauge bosons have been discussed in the past in various formulations. 
For example, a class of two-dimensional non-linear $\sigma$-models with local gauge symmetries allows one to construct composite gauge bosons \cite{Lue,Ei}. 
For various different other approaches see \cite{CGB1,CGB2,CGB3,CGB4,CGB5,CGB6,Fer,CGB7}. 
In particular, within a Hamiltonian formulation of quantum link models the link operators have been expressed as bilinears of creation and annihilation operators for rishons \cite{WR1,WR2}, showing some analogies to our setting. 
The implementation of local gauge symmetries on the level of ``constituents'' can be done easily by the use of local invariants. 
What is less clear is the realization of non-abelian gauge models that lead to asymptotic freedom and confinement as in the usual setting for quantum chromodynamics (QCD). 
The present paper proposes such models in the context of a well defined regularized functional integral for constituents.

A suitable setting for regularized gauge theories are lattice theories. 
We will follow this road and formulate our models on an euclidean space-time lattice. 
In contrast to Wilson's original proposal \cite{Wi} we will not use link variables as basic degrees of freedom. 
We rather want to know if it is possible to formulate gauge theories in terms of fundamental fermions or scalars $\psi(x)$ which are associated to each lattice point $x$, 
and transform homogeneously under local gauge transformations as $\psi(x)\to V(x)\psi(x),V^\dagger V=1$. 
For the case of fermions the presence of local gauge symmetries has already been observed in a discussion of lattice spinor gravity \cite{LSG}. 

Our approach can also be linked to models with an emergent metric. 
Any higher-dimensional model which exhibits diffeomorphism symmetry and a collective metric or vielbein can yield four-dimensional gauge theories \cite{CWD16}. 
They arise after dimensional reduction for an arbitrary ``internal'' space with isometries \cite{K1,K2,CWSO10,Wein}.

The formulation of an action with local gauge symmetry is rather straightforward by employing local invariants formed from $\psi(x)$. 
However, it is not clear if such generalized gauge theories based on scalars or fermions will produce the striking dynamical features of asymptotic freedom and confinement that characterize standard non-abelian gauge theories. 
In this paper we argue that this is indeed possible for a suitable region in the space of parameters characterizing the gauge invariant action. 
This opens the road for constructing the standard model of particles in terms of other ``fundamental'' degrees of freedom, for example purely in terms of fermions. 
Furthermore, gauge theories for ``fundamental'' scalars or fermions may offer interesting perspectives for a realization of $d$-dimensional gauge theories by ultracold atoms \cite{CWi,Ho,Or,Bu,Ba1,Ba2}.

In this paper we concentrate on ``fundamental'' scalar fields rather than on fermions. 
The reason for this is an easy access to such theories by numerical simulations. 
Such simulations will be very useful to verify the suggestions made in the present work. 
We present an explicit lattice model for scalar fields for which we predict confinement. 
The generalization to ``fundamental'' fermions is briefly discussed in the conclusions. 

More in detail, we will investigate models with $NN_f$ complex scalar fields $\chi^a_i(x)$, with $i=1\dots N$ the ``gauge index'' on which $SU(N)\times U(1)$ gauge transformations act as $\chi^a_i(x)\to V_{ij}(x)\chi^a_j(x)$.
We consider $N_f$ flavors, labeled by the index $a$. Gauge invariant combinations of scalars can be formed as 
\be\label{1a}
M^{ab}(x)=\big(\chi^a_i(x)\big)^*\chi^b_i(x). 
\ee
A functional integral for which the action can be written uniquely in terms of the hermitean $N_f\times N_f$ matrices $M(x)$ is invariant under local gauge transformations without the need to introduce explicit link variables.
In order to be explicit we take four space-time dimensions and $N_f = 8N$.
For the particular example $N=3$, $N_f=24$ the model has the same symmetries as QCD coupled to 24 massive colored scalars. (Smaller values of $N_f/N$ may also be viable.)
The composite fields $M(x)$ can be associated with ``mesons'' formed from two scalars.

An example for such an action has the naive continuum limit
\ba\label{AA}
S&=&\int d^4 x\big[\bar m^2{\rm tr}M^2+\bar\lambda{\rm tr}\big\{M^2\partial_\mu M\partial_\mu M\big\}\nn\\
&&-\frac{3}{2}{\rm tr}\big\{M\partial_\mu MM\partial_\mu M\big\}\big].
\ea
This action is positive for $\bar m^2>0,\bar\lambda>3/2$, with minimum for $M(x)=0$. 
It is characterized by a derivative interaction involving four powers of $M(x)$ or eight powers of $\chi(x)$. 
Besides local $SU(N)\times U(1)$ gauge symmetry and space-time symmetries it is also invariant under global $SU(N_f)$-flavor transformations acting on the flavor indices of $\chi$ or $M$. 
A precise lattice formulation of the action will be given in the next section and is important for the detailed understanding. 
Extensions of this action will be discussed in later parts of this paper. 
(For example, a kinetic term $\sim \tr \{\partial_\mu M\partial_\mu M\}$ could be added without changing the qualitative features.)
While the action \eqref{AA} only involves mesons, the basic functional integral is formulated in terms of the ``colored'' scalars $\chi$.
We will see that the action can be reformulated in terms of other collective fields as gauge bosons.

Due to the complicated structure of the interaction in eq. \eqref{AA} it is not easy to guess the dynamics of the low momentum excitations of this model. 
We will argue that for suitable $\bar m^2$ and $\bar \lambda$ it actually belongs to the same universality class as a pure $SU(N)\times U(1)$ gauge theory. 
While the naive continuum limit \eqref{AA} may provide for a reasonable description at microscopic distances only somewhat larger than the lattice distance, 
the true continuum limit is then better described by a standard non-abelian gauge theory with collective gauge bosons. 
In order to show this we will use collective link variables and perform a Hubbard-Stratonovich-type transformation \cite{H,S} of the functional integral. 

In a particular limit we will demonstrate the equivalence of scalar lattice gauge theory and standard lattice gauge theory based on unitary link variables. 
For this purpose we choose $\bar m^2=9\bar\lambda l^2_0 a^{-3}$, with $a$ the lattice distance.
One finds that in the limit $\bar\lambda\to\infty$, $l^2_0$ fixed, only unitary collective link variables determine the long range degrees of freedom. 
This limit realizes a standard lattice gauge theory with microscopic gauge coupling determined by $g^2=6/\beta=2l^{-4}_0$. 
For large $l_0$ one expects for intermediate distances a weak coupling regime characterized by a running gauge coupling realizing asymptotic freedom, 
while a strong coupling regime for large distances is characterized by confinement and glueballs. 
Similar features are expected to hold for finite large $\bar\lambda$. 

In order to make contact with the standard continuum formulation of gauge theories one may introduce the scalar and vector bilinears
\ba\label{AAB}
\tilde S_{ij}(x)&=&\chi^a_i(x)\big(\chi^a_j(x)\big)^*,\nn\\
(\tilde B_\mu)_{ij}(x)&=&-i\chi^a_i(x)\big(\partial_\mu\chi^a_j(x)\big)^*.
\ea
For an invertible matrix $\tilde S$ one can define collective gauge fields (in a matrix notation)
\be\label{AAC}
\tilde A_\mu(x)=\frac12\big(\tilde S^{-1}(x)\tilde B_\mu(x)+\tilde B^\dagger_\mu(x)\tilde S^{-1}(x)\big).
\ee
With respect to local gauge transformations they have the standard inhomogeneous transformation properties. 
With $\tr \tilde S=\Sigma_aM_{aa}$ one associates $l_0$ with a constant value of $\tr\tilde S$, $l_0=\tr \tilde S/N$. 
One can re-express the action in terms of the collective scalars $\tilde S$ and $M$ and the collective gauge fields $\tilde A_\mu$ or, alternatively, in terms of $\chi$ and $\tilde A_\mu$. 
The collective fields \eqref{AAB}, \eqref{AAC} are the basis for a Hubbard-Stratonovich transformation and a formulation of the continuum limit for the effective action in terms of gauge fields. 

On the other side, we will show that the functional integral can be reformulated in terms of $M(x)$ instead of $\chi(x)$. 
This eliminates any direct appearance of the local gauge transformations since $M(x)$ is invariant. 
The formulation in terms of $M$ involves a Jacobian arising from the transition from $\chi$ to $M$ in the functional measure. 
This adds an effective non-polynomial potential term to the action \eqref{AA}. 
In this ``meson formulation'' the minimum of the action occurs for $\kl M\kr\neq 0$. 

One and the same theory is therefore formulated in terms of different degrees of freedom: 
colored scalars $\chi$, invariant scalars $M$, or gauge fields or associated link variables coupled to colored scalars. 
Understanding the connections between the equivalent formulations may be useful for the understanding of the structure of gauge theories, in particular if fermions (quarks) are added. 

This paper is organized as follows: In sect. \ref{Invariant action and} we discuss the formulation of a lattice theory with local gauge symmetry in terms of colored scalars $\chi$. 
We introduce scalar bilinears that transform as link variables and partly reformulate the action in terms of those link bilinears. 
In contrast to the usual link variables in lattice gauge theories the link bilinears are not constrained to be unitary matrices - they can take arbitrary values. 

In sect. \ref{Gauge bosons as collective excitations}  we employ a formalism with collective link variables and perform a transformation of the functional integral to a formulation with link variables and colored scalar fields. 
Even though the interactions between the scalars and gauge bosons (link variables) are rather complicated, 
the basic structure of the combined action for link variables and scalars is rather simple.
The link variables are general complex $N\times N$ matrices.
We decompose the linear link variables into unitary matrices which describe the gauge bosons and other fields.
These other fields can be very heavy such that they do not contribute in the continuum limit.
Also the colored scalars correspond after the transformation to massive scalar fields with comparatively simple self-interactions.
Then only unitary link variables play a role in the continuum limit.
The action for the unitary link variables is the standard Wilson plaquette action.
This suggests that the model is in the same universality class as standard lattice gauge theory. 
In sect. \ref{Phase diagram of scalar} we propose explicitly a choice of parameters for which we expect that the model realizes a standard weak coupling lattice gauge theory, 
with confinement scale far below the lattice cutoff.
The complicated interactions between link variables and colored scalars are subdominant in this case.
In particular, we argue that a suitably taken limit $\bar\lambda\to\infty$ is equivalent to standard lattice gauge theory. 

In sect. \ref{Gauge invariant formulation} we turn to a pure singlet formulation of scalar lattice gauge theory in terms of the ``composite'' scalar fields or ``mesons'' $M(x)$.
This realizes a formulation of gauge theories only in terms of gauge invariant quantities. 
However, the price to pay is a rather complicated non-polynomial potential for the mesons.
Our conclusions are drawn in sect. \ref{Conclusions}. 

\section{Invariant lattice action and collective link variables}
\label{Invariant action and}

\noindent {\bf 1. Scalar lattice action}

We start with $N\times N_f$ dimensionless complex scalar fields $\chi^a_i(x)$. 
Here $i=1\dots N$ is a color index, and $a=1\dots N_f$ a flavor index. 
The coordinates $x^\mu$ denote points of a $d$-dimensional hypercubic lattice, $x^\mu=an^\mu$, with integer $n^\mu$ and $a$ the lattice distance. 
Periodic or other boundary conditions may be imposed such that the number of lattice points ${\cal N}$ is finite, with continuum limit ${\cal N}\to\infty$ taken at the end. 
The local gauge symmetries act on the color index, $\chi^a_i(x)'=V_{ij}(x)\chi^a_j(x)$, $V^\dagger V=1$.

For a functional integral 
\be\label{1A}
Z=\int {\cal D}\rchi e^{-S[\rchi]}=\left( \prod_x\prod_i\prod_a \int d\rchi^a_i(x) d\rchi^a_i(x)^*\right) e^{-S[\rchi]}  
\ee
the dynamics of the model is determined by the form of the microscopic action $S$.
We will concentrate on an action that can be written in terms of the meson bilinears
\be\label{1Aa}
M^{ab}(x)=\big(\chi^a_i(x)\big)^*\chi^b_i(x),
\ee
as well as lattice derivatives thereof. Here we define lattice derivatives by
\be\label{5}
\partial_\mu f(x)=\big(f(x+e_\mu)-f(x)\big)/a, 
\ee
with $e_\mu$ a lattice unit vector in the $\mu$-direction. We will represent $M^{ab}(x)$ as an $N_f\times N_f$ matrix $M(x)$. 
Eq. \eqref{1Aa} involves a sum over the color indices $i$. 
Thus $M(x)$ is invariant under local gauge transformations, guaranteeing the gauge invariance of the action. 
Generalizations and details of possible actions with local gauge symmetry are discussed in appendix A. 
We concentrate on $N_f= 2dN$, for reasons discussed in appendix B.
Our particular example has $d=4$, $N=3$, $N_f=24$.
For $N_f>N$ one has restrictions on the allowed values of $M$ that can be obtained from eq. \eqref{1Aa}, as $\det M=0$. 
This does not harm since the basic degrees of freedom are the scalars $\chi$. 
(For $N_f=N$ both $\chi$ and $M$ are quadratic $N\times N$ matrices. 
Again, there are some positivity restrictions on $M$.) 

We concentrate on the gauge invariant scalar lattice action
\ba\label{30C}
S&=&\bar S_p+\bar S_l,\nn\\
\bar S_p&=&\sum_{\rm plaquettes}\bar{\cal S}_p~,~\bar S_l=\sum_{\rm links}\bar{\cal S}_l,
\ea
where the sum over links is $\sum_x\sum_\mu$, while the sum over plaquettes corresponds to $\sum_x\sum_\nu\sum_{\mu<\nu}$. 
The ``scalar link action'' $\bar\s_l$ relates to a link $\m$ joining the points $x$ and $x+e_\mu$ and reads
\ba\label{30D}
\bar{\cal S}_l&=&\frac14\left(\bar\lambda-\frac{d-1}{2}\right){\rm tr}
\Big\{\big[M(x+e_\mu)+M(x)\big]^2\nn\\
&&\quad\times [M(x+e_\mu)-M(x)]^2\Big\}\nn\\
&&+\frac{\bar m^2}{4d}a^{\frac{d+2}{2}}{\rm tr}\big\{[M(x+e_\mu)+M(x)]^2\big\}.
\ea
For $\bar \lambda \geq (d-1)/2$ and $\bar m^2\geq 0$ it is positive definite. 

The ``scalar plaquette action'' $\bar\s_p$ involves the matrices $M(x)=M_1,M(x+e_\mu)=M_2,M(x+e_\nu)=M_3$ and $M(x+e_\mu+e_\nu)=M_4$ for the four points 
$(x,x+e_\mu,x+e_\nu,x+e_\mu+e_\nu)$ belonging to a plaquette $(x;\mu\nu)$. 
It is defined as 
\ba\label{M1}
\bar\s_p&=&\frac18{\rm tr}\left\{(M^2_1+M^2_4)(M_3-M_2)^2 \right. \nn\\
&&+(M^2_2+M^2_3)(M_4-M_1)^2\nn\\
&&-2M_1(M_3-M_2)M_4(M_3-M_2)\nn\\
&&\left. -2M_2(M_4-M_1)M_3(M_4-M_1)\right\}.
\ea
In appendix A we show $\bar\s_p\geq 0$. 
There we also establish that the lattice action \eqref{30C} has (for  $d=4$) the  naive continuum limit \eqref{AA}. 
The scalar lattice action \eqref{30C} can therefore be viewed as the lattice regularization of the action \eqref{AA}. 

The action \eqref{30C} is invariant under lattice translations and $\pi/2$-rotations. 
A reflection of the coordinate $x^\mu$ exchanges $M_1\leftrightarrow M_2,M_3\leftrightarrow M_4$, in addition to a change of position of plaquettes and links. 
The action is invariant under this transformation. 
This also holds for a diagonal reflection $x^\mu\leftrightarrow x^\nu$, which is associated to $M_2\leftrightarrow  M_3$. 
Charge conjugation $\chi\to\chi^*$ results in $M\to M^T$ and leaves $S$ invariant. 
Coordinate reflections and charge conjugation change the order of matrices in $\bar\s_p$, e.g. $\tr \{M_1M_2M_4M_3\}\leftrightarrow \tr \{M_1M_3M_4M_2\}$.

As a functional of arbitrary $M(x)$ the action \eqref{30C} would have a minimum for any configuration where $M(x)=M_0$ for $x$ even, 
and $M(x)=-M_0$ for $x$ odd, where for even $x$ one has $\sum\nolimits_\mu n^\mu=$ even, while for odd $x$ the sum of the $d$ integers $n^\mu$ is odd. 
Two neighboring sites belonging to a link $\m$ have then opposite $M_0$ such that $\bar\s_l$ vanishes. 
Furthermore, one concludes from $M_4=M_1, M_3=M_2$ that $\bar\s_p$ is zero as well. 
However, $M$ is a composite \eqref{1Aa} of the scalars $\chi$ and cannot take arbitrary values. 
For example, the diagonal elements obey $M^{aa}\geq 0$. 
The only way to realize both matrices $M_0$ and $-M_0$ is for $\chi_0=0,M_0=0$. 
This defines the minimum of the action or the ground state. 
In appendix A we discuss possible generalizations of eq. \eqref{1Aa} for which arbitrary values of $M(x)$ can be realized.

While for our setting with $M(x)$ defined by eq. \eqref{1Aa} the ground state $M=0$ is simple, the understanding of the fluctuations is a rather complicated task due to the complex structure of the derivative terms. 
The issue will become more clear if we discuss $\bar\s_p$ and $\bar\s_l$ in some more detail. 

The scalar lattice action \eqref{30C} depends on two parameters, $\bar\lambda$ and $\bar m^2$. 
(One can scale the coefficient of $\bar S_p$ to one by a suitable multiplicative rescaling of $\chi$ and therefore $M$.) 
The main result of this paper is the suggestion that for suitable values of $\bar\lambda$ and $\bar m^2$ this model belongs to the same universality class as standard gauge theories, 
showing asymptotic freedom and confinement in the non-abelian sector. 
In particular, we argue that for large $\bar\lambda$ and $\bar m^2$, with $\bar\lambda/\bar m^2$ fixed, our model describes a standard gauge theory with gauge coupling $g^2=6/\beta=162\bar\lambda^2 a^{-6}\bar m^{-4}$. 
It is possible to verify (or falsify) this conjecture by a numerical simulation of the scalar functional integral \eqref{1A}-\eqref{M1}. 

\medskip\noindent
{\bf 2. Collective link variables}

A central role of this work will be played by link bilinears. They involve scalars at neighboring sites, with flavor indices contracted,
\be\label{7}
\tilde L_{ij}(x;\mu)=\chi^a_i(x)\big(\chi^a_j(x+e_\mu)\big)^*.
\ee
One may consider the link bilinears as complex $N\times N$-matrices $\tilde L(x;\mu)$.
Under local unitary gauge transformations, represented by unitary $N\times N$ matrices $V(x)$, the links transform as
\be\label{10}
\tilde L(x;\mu)\to V(x)\tilde L(x;\mu)V^\dagger(x+e_\mu),
\ee
similar to the links in lattice gauge theories. 
The link bilinears show analogies to the construction of quantum link operators from rishons within a Hamiltonian formalism \cite{WR1,WR2}.
We will in the following use a notation for $\chi$ as an $N\times N_f$ matrix, where the first index is the color index and the second the flavor index, $\chi_{ia}(x)\equiv \chi^a_i(x)$. 
Correspondingly, the $N\times N$ matrices $\tilde L(x,\mu)$ and the $N_f\times N_f$ matrices $M(x)$ obey
\be\label{13B}
\tilde L(x;\mu)=\chi(x)\chi^\dagger(x+e_\mu)~,~M(x)=\chi^\dagger(x)\chi(x).
\ee

By use of the transformation property \eqref{10} one can use the link bilinears for the construction of gauge invariant quantities similar to Wegner-Wilson loops or Polyakov loops. 
For example, two static sources representing infinitely heavy quarks can be joined by a chain of links in order to make the configuration gauge invariant. 
This leads to the usual confinement criteria, now formulated in terms of loops formed from the collective link variables. 
We observe that all closed loops can be expressed as suitable chains of ``meson bilinears'' $M$. 

Local invariants can be constructed from link bilinears in a standard way. 
A plaquette $(x;\mu\nu)$ can be used for defining the invariant
\ba\label{11}
\tilde P(x;\mu,\nu)&=&{\rm tr}\big\{\tilde L(x;\mu)\tilde L(x+e_\mu;\nu)\nn\\
&&\times\tilde L^\dagger(x+e_\nu;\mu)\tilde L^\dagger(x;\nu)\big\}.
\ea
Defining links with negative directions as 
\be\label{12}
\tilde L(x+e_\mu;-\mu)=\tilde L^\dagger(x;\mu),
\ee
the invariant $\tilde P$ corresponds to the trace of the product of four link matrices around the plaquette. 
The notation $\tilde P(x;\mu,\nu)$ denotes a start at $x$, first link in direction $\mu$, second in direction $\nu$, third direction $-\mu$, and fourth direction $-\nu$. 
We observe that, in general, $\tilde P$ has an orientation,
\be\label{13}
\tilde P(x;\nu,\mu)=\tilde P^*(x;\mu,\nu).
\ee
We will see below that the plaquette invariants $\tilde P$ play an important role for the understanding of the plaquette action $\s_p$ and for the close relation of our model to standard lattice gauge theories.
Using the definitions \eqref{1Aa}, \eqref{7}, we can write $\tilde P$ as a product of meson invariants,
\be\label{13A}
\tilde P(x;\mu,\nu)={\rm tr}\big\{M(x)M(x+e_\mu)M(x+e_\mu+e_\nu)M(x+e_\nu)\big\}.
\ee

An interesting subclass of configurations $\chi$ are those for which all link bilinears $\tilde L\m$ are ``pseudo-unitary'' of the form
\be\label{15A1}
\tilde L\m = \tilde l_0 \tilde U\m ,
\ee
with constant $\tilde l_0$ and $\tilde U\m$ unitary matrices.
For the subclass of those configurations $\chi$ the invariant $\tilde P$ looks already rather close to the plaquette action for standard lattice gauge theories with unitary link variables.
For models where arbitrary unitary matrices $\tilde U\m$ can be realized for suitable $\chi$ 
it seems likely that there exist possible choices of an action such that the model is in the same universality class as standard lattice gauge theories.
For our example with $N_f = 2dN$ the number of degrees of freedom contained in $\chi$ is twice the number of degrees of freedom in $\tilde L$.
It seems therefore possible (cf. appendix B) to find for any arbitrary configuration of links $\tilde L$ a set of suitable scalar configurations $\chi$ for which eq. \eqref{7} holds.
In particular, this can realize arbitrary $\tilde U$ in eq. \eqref{15A1}.

On the other hand, among the configurations $\chi$ obeying eq. \eqref{15A1} 
we may call those ``trivial'' for which all pseudo-unitary link bilinears can be simultaneously transformed by suitable gauge transformations \eqref{10} 
to a constant matrix proportional to the unit matrix, $\tilde L\m=\tilde l_0$.
If pseudo-unitary link bilinears do not exist or are all realized by trivial configurations $\chi$ it seems likely that the model belongs to a universality class which differs from standard lattice gauge theories.
Such a universality class exhibits still a local gauge symmetry, but we do not expect the characteristic features of asymptotic freedom and confinement.

For $N_f < N$ the determinant of the link bilinear \eqref{7} vanishes. No pseudo-unitary link bilinears exist in this case.
This shows the general tendency that for small $N_f / N$ confinement may not be realized even for non-abelian gauge symmetries with $N>1$.
In contrast, large enough $N_f/N$ are expected to yield the standard setting with confinement. 
The issue may also be demonstrated with the simplest example $N_f=1$.
For only one flavor we would have $M(x)=M^*(x)$ such that $\tilde P(x;\mu,\nu)=\tilde P^*(x;\mu,\nu)$, cf. eq. \eqref{13A}. 
Then $\tilde P$ is blind to the angles in the decomposition $\chi_i(x)=\hat U_{i1}(x)k(x),\,k(x)\in{\mathbbm R},\,\hat U^\dagger\hat U=1$, since $M(x)=k^2(x)$ does not involve $\hat U$. 
Such models cannot reproduce the angular dependencies of the plaquette invariant in standard lattice gauge theories. 
We will discuss in the appendix B for which choice of the number of flavors $N_f$ one may expect to find the same universality class as for a standard lattice gauge theory. 
Typically, this requires a minimal $N_f$ for a given $N$.
For our choice of $N_f = 2dN$ we argue that the action \eqref{30C} with large $\bar \lambda$ indeed realizes the same continuum limit as a standard lattice gauge theory.
We will show this in the next section by means of a Hubbard-Stratonovich transformation.

For large enough $N_f/N$ (as for our example $N_f = 2dN$) the link bilinears $\tilde L$ are arbitrary complex $N \times N$ matrices. 
While the collective gauge bosons are described by pseudo-unitary link bilinears \eqref{15A1}, the more general link variables also contain other fields.
In appendix C we discuss the field content of the link variables $\tilde L\m$.
This also gives an explicit expression for the collective gauge bosons.
 
\medskip\noindent
{\bf 3. Action in terms of link bilinears}

We can partly reformulate the action \eqref{30C} in terms of the collective link bilinears. 
For this purpose we employ a type of kinetic term for the link variables 
\ba\label{23FA}
{\cal L}_p(x;\mu,\nu)&=&\frac14{\rm tr}\left\{\tilde H^\dagger_{\mu\nu}(x)\tilde H_{\mu\nu}(x)\right.\\
&&\left. +\tilde H^\dagger_{\nu-\mu}(x+e_\mu)\tilde H_{\nu-\mu}(x+e_\mu)\right\},\nn
\ea
with 
\ba\label{23FB}
\tilde H_{\mu\nu}(x)&=&\tilde L\m\tilde L(x+e_\mu;\nu)\nn\\
&&-\tilde L(x;\nu)\tilde L(x+e_\nu;\mu),\nn\\
\tilde H_{\nu-\mu}(x+e_\mu)&=&\tilde L(x+e_\mu;\nu)\tilde L^\dagger(x+e_\nu;\mu)\\
&&-\tilde L^\dagger\m L(\tilde x;\nu)\nn.
\ea
We observe that $\tilde H_{\nu-\mu}(x+e_\mu)$ obtains from $\tilde H_{\mu\nu}(x)$ by a $\pi/2$-rotation. 
Summing ${\cal L}_p$ over all plaquettes preserves the lattice rotation symmetries. 
The form of eq. \eqref{23FA} guarantees invariance under local gauge transformations and establishes ${\cal L}_p\geq 0$. 
We can write ${\cal L}_p$ as a sum of invariants $\sim\tilde P$ plus other terms, as explained in more detail in appendix A.

One can express ${\cal L}_p$ in terms of the scalar bilinears $M$
\ba\label{M2}
{\cal L}_p&=&\frac14\tr \big\{
M_1(M_3-M_2)M_4(M_3-M_2)\nn\\
&&+M_2(M_4-M_1)M_3(M_4-M_1)\big\}.
\ea

This makes the gauge invariance manifest.
Comparing with eq. \eqref{13A} the terms in eq. \eqref{M2} with one matrix at each site can be identified with $\tilde P$ or $\tilde P^*$.
Additional terms involve one given matrix twice, as $\tr\{M_1M_3M_4M_3\}$. 
One can verify easily that the expression \eqref{M2} is invariant under lattice rotations by $\pi/2$ and lattice reflections. 
We associate ${\cal L}_p$ with the ``link plaquette action'' and define for later purposes
\be\label{16}
S_p=\sum_{{\rm plaquettes}}{\cal L}_p(x;\mu,\nu).
\ee
This is different from the sum of scalar plaquette actions $\bar S_p=A_p-S_p$ in eq. \eqref{30C}. 

Indeed, we can write the scalar plaquette action $\bar\s_p$ as a difference of two positive semidefinite terms
\be\label{M3}
\bar\s_p={\cal A}_p-{\cal L}_p.
\ee
Here
\be\label{M4}
{\cal A}_p=\frac18{\rm tr}\big\{(M^2_1+M^2_4)(M_3-M_2)^2+(M^2_2+M^2_3)(M_4-M_1)^2\big\}
\ee
has a comparatively simple structure. It has the tendency to suppress inhomogeneities in the configuration $M(x)$. 
However, it competes with the link plaquette term $\sim {\cal L}_p$ which has a negative sign. 
The most difficult part in the understanding of the action \eqref{30C} is due to the part $-S_p$ which favors nonzero values for the link bilinears. 
The point $\chi=0$ has zero weight in field space such that all relevant configurations will have $\chi\neq 0$. 
For $\chi\neq 0$ the positive term ${\cal A}_p$ and the negative term $-{\cal L}_p$ tend to drive the fluctuations into opposite directions, somewhat similar to frustrated systems. 
It is this type of competition that will finally be responsible for the confinement characteristic for standard non-abelian gauge theories. 

A similar type of competition can be observed for the scalar link action \eqref{30D}. 
One can employ the identities
\be\label{23A}
\tr\big \{[\tilde L^\dagger\m\tilde L\m]^K\big \}=\tr
\big \{[M(x)M(x+e_\mu)]^K\big \}
\ee
in order to write
\be\label{23B}
\bar{\cal S}_l={\cal A}_l-{\cal L}_l,
\ee
with 
\ba\label{23C}
{\cal L}_l&=&\frac12\left(\bar\lambda-\frac{d-1}{2}\right)\tr\big\{[\tilde L^\dagger\m\tilde L\m]^2\big\}\nn\\
&&-\frac{\bar m^2}{2d}a^{\frac{d+2}{2}}\tr\big\{\tilde L^\dagger\m\tilde L\m\big\}.
\ea
(Note that matrix multiplications and traces in eq. \eqref{23A} are in different spaces.) 
The expression \eqref{23C} is a type of ``Mexican hat potential'' for the link variables. 
For large $\tilde L$ the piece $-{\cal L}_l$ is unbounded from below. 
Still, the boundedness of $\bar{\cal S}_l$ is preserved by the competing term ${\cal A}_l$,
\ba\label{26A}
\mathcal{A}_l &=& \frac14 \left(\bar\lambda - \frac{d-1}{2}\right) \tr\left\{ M^4(x+e_\mu)+M^4(x)\right\} \nn \\
&+& \frac{\bar m^2}{4d} a^{\frac{d+2}{2}} \tr\left\{ M^2(x+e_\mu) + M^2(x) \right\}.
\ea

In summary of this section we have formulated the action \eqref{30C}-\eqref{M1} of scalar lattice gauge theory.
It is invariant under local gauge symmetries. 
Parts of the action, namely ${\cal L}_p$ and ${\cal L}_l$, can be expressed in terms of collective link bilinears. 
This is not possible for ${\cal A}_p$ and ${\cal A}_l$. 
The competition between the positive contributions ${\cal A}_p$ and ${\cal A}_l$ and the negative contributions 
$-{\cal L}_p$ and $-{\cal L}_l$ seems crucial for realizing scalar models that are in the same universality class as usual gauge theories. 
Omitting ${\cal L}_p$ and ${\cal L}_l$ would still lead to an action with local gauge symmetry. 
Presumably, this belongs to a different universality class. 

\medskip\noindent
{\bf 4. Choice of action and Hubbard-Stratonovich 

\hspace{0.2cm}transformation}

At this point we should motivate the particular choice \eqref{30C} of the scalar lattice action.
The reason for the selection of this apparently rather complicated structure is the simplicity of the dominant terms after a type of Hubbard-Stratonovich transformation
which introduces explicit link variables in the functional integral.
This transformation will replace the terms $-\mathcal{L}_p[\tilde L]$ and $-\mathcal{L}_l[\tilde L]$ by terms $+\mathcal{L}_p[L]$ and $+\mathcal{L}_l[L]$, where $L$ are the explicit link variables and $\tilde L$ the link bilinears.
The opposite sign resulting from a change to explicit link variables is characteristic for this type of transformation.
If one wants to obtain an action for the explicit link variables $L$ that is well defined, one needs the positive signs of $\mathcal{L}_p[L]$ and $\mathcal{L}_l[L]$.
In turn, the negative sign of the terms $-\mathcal{L}_p[\tilde L]$ and $-\mathcal{L}_l[\tilde L]$ in the pure scalar formulation needs compensating terms in 
$\mathcal{A}_p$ and $\mathcal{A}_l$ in order to guarantee the positivity of the scalar action \eqref{30C}.

The issue may be demonstrated by a simple example. Consider a real scalar $s$ and a complex scalar $\chi$, with action $S = \int_x\mathcal{L}$,
\be\label{26B}
\mathcal{L} = \frac12 \partial_\mu s\, \partial_\mu s + \frac12 m^2 s^2 +\partial^2 s\, \chi\chi^* - m^2s\,\chi\chi^* + \mathcal{L}_\chi,
\ee
with $\mathcal{L}_\chi$ depending only on $\chi$ and not on $s$.
The Gaussian functional integral over $s$ can be performed, resulting in an action involving only $\chi$,
\ba\label{26C}
\mathcal{L}^{(\chi)} &=& \mathcal{L}_\chi - \frac12 \partial_\mu \tilde s \, \partial_\mu \tilde s - \frac12 m^2 \tilde s^2, \nn \\
\tilde s &=& \chi\chi^*.
\ea
We observe the negative sign of the terms involving the bilinear $\tilde s = \chi\chi^*$.
Inversely, we could start with a model involving only the complex scalar $\chi$ and action given by $\mathcal{L}^{(\chi)}$ in eq. \eqref{26C}.
This can be transformed by a Hubbard-Stratonovich transformation \cite{H,S} to the coupled system of $s$ and $\chi$ with action \eqref{26B}.

In our setting, the role of $-\partial_\mu \tilde s \, \partial_\mu \tilde s$ is played by $-\mathcal{L}_p$, and $-m^2\tilde s^2$ corresponds to $-\mathcal{L}_l$.
The transformation is a bit more involved since $\mathcal{L}_p$ and $\mathcal{L}_l$ are quartic in $\tilde L$ instead of quadratic.
We will give the details of the transformation to explicit link variables in the next section.

\section{Gauge bosons as collective excitations}
\label{Gauge bosons as collective excitations}

In this section we discuss the effective action for collective gauge bosons.
In our lattice formulation this corresponds to the effective action for the link bilinears.
We will introduce a formulation in terms of explicit link variables. 
Indeed, by means of a Hubbard-Stratonovich type transformation 
we will discuss a functional integral for scalars and link variables that is equivalent to the scalar lattice gauge theory presented in sect. \ref{Invariant action and}.
In this formulation, we are already closer to the standard formulation of lattice gauge theories.
Besides the additional coupling to the colored scalars $\chi$ there remains an important difference, however.
The link variables are arbitrary complex $N \times N$ matrices instead of the unitary matrices in the standard formulation.
We deal with the corresponding ``linear lattice gauge theory in a separate publication \cite{CWto}.
(For earlier work on this issue see \cite{NP, GM, PPV, KU}.)
Our aim is a demonstration that confinement occurs for some region in the parameter space of the action \eqref{30C}.
For this purpose it will be sufficient to consider the regime where both the colored scalars $\chi$ and the excitations which move $L$ away from a pseudo-unitary matrix $L=l_0 U$, $U^\dagger U=1$, correspond to heavy excitations.
These heavy excitations become negligible in the continuum limit.
We will therefore mainly concentrate on the effective action for the link variables resulting from the Hubbard-Stratonovich type transformation.
At the end of this section we will argue that after the transformation the remaining fluctuations of $\chi$ are suppressed, and that setting $\chi  = 0$ is a reasonable approximation.
In particular, we will argue in sect. \ref{Phase diagram of scalar} that scalar lattice gauge theory is equivalent to pure standard lattice gauge theory for an appropriate limit $\bar\lambda\to\infty$ for the action \eqref{30C}. 

\medskip\noindent
{\bf 1. Effective action for scalars and link variables}

The appropriate formalism for collective fields introduces sources $K_{ij}\m$ for the link bilinears $\tilde L_{ij}\m$ in addition to the sources $j^a_i(x)$ for the ``fundamental scalars'' $\chi^a_i(x)$,
\be\label{30}
Z[j,K]=\exp W[j,K]
=\int {\cal D}\rchi\exp \big\{-S[\rchi]-S_{jK}\big\},
\ee
with source term 
\be\label{67Aa}
S_{jK}=-\sum_x\big[(j^a_i)^*\chi^a_i+\sum_\mu K^*_{ij}\m\tilde L_{ij}\m+c.c\big].
\ee
The definition \eqref{7} for the link bilinears implies an identity for the functional dependence of $W$ on $K$ and $j$,
\ba\label{31}
&&\kl \tilde L_{ij}\m\kr=\frac{\partial W}{\partial K^*_{ij}\m}\\
&&=\frac{\partial^2 W}{\partial\big(j^a_i(x)^*\partial j^a_j(x+e_\mu)}+\frac{\partial W}{\partial\big(j^a_i(x)\big)^*}
\frac{\partial W}{\partial j^a_j(x+e_\mu)}.\nn
\ea
Here $\kl \tilde L_{ij}\m\kr$ denotes the expectation value of the link bilinear in the presence of arbitrary sources $j,K$. We define 
\be\label{32}
\varphi^a_i(x)=\kl \chi^a_i(x)\kr=\frac{\partial W}{\partial\big(j^a_i(x)\big)^*}
\ee
and (``background'') link variables
\be\label{33}
\bar L_{ij}\m=\kl\tilde L_{ij}\m\kr.
\ee

The quantum effective action $\Gamma[\varphi,\bar L]$ obtains from $W[j,K]$ by the usual Legendre transform
\ba\label{34}
\Gamma [\varphi,\bar L]&=&-W[j,K]+\sum_x
\big[{\rm tr}\big\{ j^\dagger(x)\chi(x)\big\}\nn\\
&&+\sum_\mu{\rm tr}
\big\{ K^\dagger \m \bar L\m\big\}+c.c\big],
\ea
where we use the matrix notation. 
The sources $j$ and $K$ are expressed in terms of $\varphi$ and $\bar L$ by inverting the relations \eqref{31}, \eqref{32}, resulting in
\be\label{35}
\frac{\partial\Gamma}{\partial\varphi(x)}=j^\dagger(x)~,~\frac{\partial \Gamma}{\partial\bar L(x)}=K^\dagger(x).
\ee
The identity \eqref{31} translates into a corresponding identity for $\Gamma$ that relates its dependence on $\bar L$ to its dependence on $\varphi$ \cite{CWBEA}. 
A computation of the effective action $\Gamma[\varphi,\bar L]$ would establish the connection between scalar lattice gauge theory and linear lattice gauge theory on a ``macroscopic level'' 
where fluctuation effects are all included. 

\medskip\noindent
{\bf 2. Link-scalar model}

In the following part of this section we will aim for a more microscopic relation where bilinears $\tilde L\m$ are related to fluctuating fields $L\m$. 
This results in an equivalent functional integral for link variables $L$ and scalars $\chi$.
The action for the link variables will take a rather simple form, bringing us close to the standard formulation of lattice gauge theories.
We argue that the interaction between $L$ and $\chi$ can be neglected if the colored scalars have a large mass.
The scalar part of the action will turn out to be substantially simpler than the action in sect. \ref{Invariant action and}. 
In particular, the competition between ${\cal A}_p$ and ${\cal L}_p$ or ${\cal A}_l$ and ${\cal L}_l$ in eqs. \eqref{M3}, \eqref{23B} will be absent. 
The price to pay is a higher number of degrees of freedom and a more complicated relation between the expectation values $\kl L\m\kr$ and $\kl \tilde L\m\kr$ and corresponding higher order correlation functions. 

Linear lattice gauge theory is formulated as a functional integral over link variables. 
In scalar lattice gauge theory this link-integration can be implemented via a Hubbard-Stratonovich type transformation \cite{H,S}. 
For this purpose one uses the identity
\ba\label{A1}
&&\int\limits^\infty_{-\infty}dL_{ij}\m\exp \big\{-f\big[L_{ij}\m\nn\\
&&\qquad \quad ~-\chi^a_i(x)\chi^{a*}_j(x+e_\mu)\big]\big\}=c,
\ea
which holds for arbitrary functions $f$ as long as $|L_{ij}|\to \infty$ implies $f(L_{ij})\to\infty$. 
The constant does not depend on fields or sources, as can be seen easily by a shift of the integration variable. 
A similar argument allows us to insert into the functional integral \eqref{1A} the expression
\ba\label{A2}
\int {\cal D}L\exp \big\{-\bar S_L\big[L_{ij}\m-\chi^a_i(x)\chi^{a*}_j(x+e_\mu)\big]\big\}=Z_L,\nn\\
\ea
where the functional integration $\int {\cal D}L$ corresponds to a product over all links of integrations over individual link variables. 
The field independent constant $Z_L$ amounts to an irrelevant multiplicative renormalization of $Z$, or additive renormalization of $W$ and $\Gamma$, and it may be set to unity. 
The ``link action'' $\bar S_L$ can be chosen arbitrarily as long as the integral $Z_L$ is well defined and $\bar S_L\to\infty$ for $|L_{ij}\m|\to\infty$.

The link variables $L$ have the same transformation property \eqref{10} as $\tilde L$. They are arbitrary complex $N\times N$ matrices.
The action for the link variables will be determined by the choice of $\bar S_L$ in eq. \eqref{A2}.
In principle, this is free, but we will concentrate on a particular choice $\bar S_L = S_L$ to be specified below.
The action $S_L$ is not quadratic in the fields and has a minimum for fields different from zero.
Therefore the transformation \eqref{A2} will introduce several new features as compared to the Hubbard-Stratonovich transformation. 
It will help us, however, to understand the close connection between scalar lattice gauge theory and linear lattice gauge theory. 

Using eq. \eqref{A2} we can write the partition function \eqref{30} as 
\be\label{A11a}
Z[j,K]=\int {\cal D}\rchi{\cal D} L\exp \big\{-S[\rchi,L]-S_{jK}\big\},
\ee
with 
\be\label{52A}
S[\rchi, L] = S_L [L-\tilde L] + S_\rchi,
\ee
and $S_\rchi$ corresponding to the scalar action given by eq. \eqref{30C}.
The action of this ``link-scalar model'' $S[\rchi, L]$ will contain a pure link action $S_L[L]$, an interaction between the link variables $L$ and the colored scalars $\chi$ 
resulting from mixed terms invoking both $L \text{ and } \tilde L$ in an expansion of $S_L[L-\tilde L]$, and a modification of the scalar part of the action resulting from $S_L[-\tilde L]$.
We first discuss the link action $S_L[L]$ and turn to the parts containing $\chi$ later.

\medskip\noindent
{\bf 3. Action for link variables}

Let us consider a choice of $S_L$ such that the action for the link variables reads
\be\label{17}
S_L=\sum_{{\rm links}}W_L\big(L(x;\mu)\big)+ S_p[L].
\ee
Here the plaquette action $S_p[L]$ is given by eqs. \eqref{16}, \eqref{23FA}, \eqref{23FB} with $\tilde L$ replaced by $L$. 
It appears in $S_L$ with a positive sign, in contrast to the negative sign in $\bar S_p=A_p-S_p$. 
The ``link potential'' $W_L$ depends only on the matrix $L$ for one given link position $(x;\mu)$. 
We will use 
\ba\label{18}
W_L(L)&=&-\mu^2\rho+\frac{\lambda N}{2}\tr(L^\dagger L)^2 \nn\\
&=& -\mu^2\rho + \frac{\lambda}{2}\rho^2 + \lambda\tau_2\:, \nn \\
\rho&=&{\rm tr}(L^\dagger L)~,~\tau_2=\frac{N}{2}{\rm tr}(L^\dagger L-\frac1N\rho)^2,
\ea
where we assume $\lambda>0$ such that $S_L$ is bounded from below, in distinction to the negative $-{\cal L}_l$ in eq. \eqref{23B}. 
(Higher order terms could be added, if necessary.)
The parameters $\mu^2$ and $\lambda$ are related to the parameters $\bar\lambda$ and $\bar m^2$ in the action \eqref{30C} by 
\be\label{XAX}
\bar\lambda=\lambda N+\frac{d-1}{2}
\ee
and 
\be\label{XAY}
\bar m^2=2d\mu^2 a^{-\frac{d+2}{2}}.
\ee

A functional integral over link variables $L$ with an action of the type \eqref{17} defines a model of ``linear lattice gauge theory''. 
In contrast to the more standard ``non-linear lattice gauge theory'', where the matrices $L$ are replaced by unitary matrices $U$ which obey the constraint $U^\dagger U=1$, the matrices $L$ are unconstrained. 
Standard lattice gauge theories with Wilson action are recovered if one replaces $L$ by $l_0 U$ with constant $l_0$. We note that $W_L$ becomes an irrelevant constant in this case.

The relation between linear and non-linear lattice gauge theories is similar to the relation between linear and non-linear $\sigma$-models \cite{JW}.
Consider a potential $W_L(L)$ that takes its minimum for a unit matrix, $L_0=l_0{\mathbbm 1}$, with real $l_0$. 
This can be realized for positive $\mu^2$ and $\lambda$, with $\rho_0=N l^2_0=\mu^2/\lambda$. 
We will next show that excitations around such a minimum describe a standard lattice gauge theory with unitary link variables coupled to additional fields in the singlet and adjoint representations of the gauge group. 

\medskip\noindent
{\bf 4. Unitary link variables and ``link scalars''}

We next establish the relation between linear lattice gauge theories and the usual lattice gauge theories based on unitary link variables. 
One can represent a complex $N\times N$ matrix $L$ as a product of a hermitean matrix $S$ and a unitary matrix (polar decomposition \cite{Dr})
\be\label{19}
L(x;\mu)=S(x;\mu)U(x;\mu)~,~
S^\dagger=S~,~U^\dagger U=1.
\ee
The gauge transformation property
\ba\label{20}
S'(x;\mu)&=&V(x)S(x;\mu)V^\dagger(x),\nn\\
U'(x;\mu)&=&V(x)U(x;\mu)V^\dagger(x+e_\mu),
\ea
implies for $U(x;\mu)$ the same transformation property as for $L(x;\mu)$, while $S(x;\mu)$ involves only the gauge transformations at $x$. 
The fields
\ba\label{21}
S\m&=&l\m+A_S\m,\nn\\
l&=&\frac1N{\rm tr} S~,~{\rm tr} A_S=0~,~A_S^\dagger=A_S,
\ea
decompose into a singlet $l\m$ and an adjoint representation $A_S\m$. 
The singlet is invariant, while $A_S$ transforms homogeneously with respect to local gauge transformations at the point $x$. 

For each site $x$ we have $d$ fields $S\m$, one for each value of the index $\mu$. 
For simplicity, we identify these fields here, $S\m=S(x;\nu)=S(x)$, where $S(x)$ is associated with a scalar field. 
These scalars may be called ``link scalars'', in distinction to the ``site scalars'' $\chi$ introduced in sect. \ref{Invariant action and}. 
While site scalars are in the fundamental representation of the gauge group, the link scalars belong to the adjoint representation or are singlets. 
A more detailed discussion of the fields contained in $S\m$ is analogous to the one in appendix C, see also ref. \cite{CWto}.

The matrices $U\m$ play the role of unitary link variables which are familiar in lattice gauge theories. 
They are related to the gauge fields $A_\mu$ (represented here as hermitean $N\times N$)-matrices) by
\be\label{28F1}
L(x;\mu)=S(x)U\m~,~U\m=\exp\big\{iaA_\mu (x)\big\}.
\ee
Infinitesimal gauge transformations of $A_\mu$ involve the usual inhomogeneous term. 
Indeed, with $V(x)=\exp\big(i\alpha(x)\big)=1+i\alpha(x),\alpha^\dagger(x)=\alpha(x)$, eq. \eqref{20} implies
\be\label{28F2}
\delta A_\mu=i[\alpha,A_\mu]-\partial_\mu\alpha.
\ee

With $LL^\dagger=SS^\dagger$ the link potential is independent of $U$, i.e. $W_L\big (L\m\big)=W_L\big(S\m\big)$.
The unitary link variables appear only in the kinetic term ${\cal L}_p$ through the invariant $\tilde P$. 
For the action \eqref{17} this implies $S_L=S_g+S_W+S_A$, with 
\ba\label{22}
&&S_g=-\sum_{{\rm plaquettes}}
\big\{ l^2(x)l(x+e_\mu)l(x+e_\nu)\operatorname{Re}\big(P_U(x;\mu,\nu)\big)\nn\\
&&\qquad -\frac N4\big[l^4(x)+l^2(x)l^2(x+e_\mu)+l^2(x)l^2(x+e_\nu)\nn\\
&&\qquad +l^2(x+e_\mu)l^2(x+e_\nu)\big]\big\}.
\ea
Here $P_U(x;\mu,\nu)$ corresponds to $\tilde P$ in eq. \eqref{11} with the replacement $\tilde L\to U$. 
For $l(x)=l_0$ the ``gauge part'' of the action $S_g$ is precisely the plaquette action of standard lattice gauge theories \cite{Wi}
\ba\label{23}
S_g&=&-\frac{\beta}{3}\sum_{{\rm plaquettes}}\big \{\operatorname{Re} P_U(x;\mu,\nu)-N\big\}~,\nn\\
\frac\beta3&=&\frac{2a^{d-4}}{g^2}= l^4_0.
\ea
In particular, for $d=4$ the gauge coupling $g$ is given by
\be\label{35A}
g^2 = \frac{2}{l_0^4}.
\ee

In addition, $S_g$ contains derivative terms for the scalar singlet $l(x)$, which read in lowest order $a^2$
\be\label{SK1}
S^{(l)}_{kin}=\frac12 N(d-1)a^2\sum_x\sum_\mu l^2(\partial_\mu l)^2+\dots
\ee

The potential part 
\be\label{37A}
S_W=d\sum_xW_L\big[l(x)+A_S(x)\big]
\ee
involves the scalar fields $l$ and $A_S$. Finally, the part $S_A$ contains covariant kinetic terms for the adjoint scalar $A$. It arises from $S_p$ and vanishes for $A=0$. This part can be found in appendix C. 

We conclude that for arbitrary complex $L$ and gauge group $SU(N)\times U(1)$ the action of linear lattice gauge theory describes gauge fields as well as scalars in the adjoint and singlet representations. Similarly, for real $L$ and gauge group $SO(N)$ the matrices $U$ are orthogonal, $U^TU=1$, and $A$ corresponds to a traceless symmetric tensor representation. 

\medskip\noindent
{\bf 5. Limit of standard non-linear lattice gauge theory}

We next show that the limit $\lambda\to\infty$ (fixed $l^2_0$) of linear lattice gauge theory results in the standard lattice gauge theory with unitary link variables. 
For this purpose we choose a potential $W_L(S)$ for which a quadratic expansion around the minimum at $S=l_0$,
\ba\label{24}
W_L(S)&=&W_0+\frac12\bar m^2_l l^2_0(l-l_0)^2+\frac12\bar m^2_Al^2_0{\rm tr}(A^2_S)+\dots,\nn\\
\bar m^2_l&=&4N^2\lambda~,~\bar m^2_A=4N\lambda,
\ea
is characterized by large positive values $\bar m^2_l\gg 1,\bar m^2_A\gg1$. Comparing with typical kinetic terms in the action 
\ba\label{25}
S^{(l,A)}_{{\rm kin}}&=&\sum_x\frac12 Z_ll^2_0a^2\partial_\mu l(x)\partial_\mu l(x)\nn\\
&&+\frac12 Z_A l^2_0a^2{\rm tr}
\big\{\partial_\mu A_S(x)\partial_\mu A_S(x)\big\}
\ea
the normalized mass terms read in the continuum limit $m^2_l=\bar m^2_l/(Z_la^2),m^2_A=\bar m^2_A/(Z_Aa^2)$.
With eq. \eqref{SK1} one has $Z_l=N(d-1)$ and 
\be\label{SK2}
m^2_l=\frac{4N\lambda}{(d-1)a^2}.
\ee
For very large $m^2_l$ and $m^2_A$ the fluctuations of the scalar fields are strongly suppressed and give only minor corrections to the functional integral. 
In the limit $\bar m^2_l\to\infty,\bar m ^2_A\to\infty$ we approximate $S(x)$ by $l_0$. 
Then only $U\m$ remains as effective degree of freedom and we expect linear lattice gauge theory to give the same results as non-linear lattice gauge theory for the corresponding value of $\beta=3l^4_0$. 
This extends to the more complicated structure of fields $S\m$. 

We conclude that our model of linear lattice gauge theory has a simple limit. 
For $\lambda\to \infty$, $\mu^2=N\lambda l^2_0\to\infty$, with fixed $l^2_0$, the linear lattice gauge theory is equivalent to the standard (non-linear) lattice gauge theory with $\beta=3l^4_0$. 
Indeed, the limit $\lambda\to\infty,\:\mu^2=N\lambda l^2_0$ can be interpreted as two constraints on the link variables
\be\label{30Aa}
\tr \{L^\dagger L\}=\tr S^2=N l^2_0,
\ee
and
\be\label{30AB}
\tr \big\{(L^\dagger L)-l^2_0)^2\big\}=\tr\big\{ (S^2-l^2_0)^2\big\}=0. 
\ee
The solution of these two constraints, $S^2=l^2_0$, fixes $S$ to be of the form $S=l_0\tilde U,\:\tilde U^\dagger \tilde U=1,\:\tilde U^\dagger=\tilde U$. 
In turn, this implies the constraint that the link variables are unitary up to an overall constant, $L=l_0U$, such that we recover a standard $SU(N)\times U(1)$ lattice gauge theory. 

Starting from the limit $\lambda\to\infty$ we may lower the values of the coupling $\lambda$ while keeping $l^2_0$ fixed. 
For smaller values of $\bar m^2_l$ and $\bar m^2_A$ we still expect the model to be in the same universality class as standard lattice gauge theories.
The long distance behavior will be characterized by the value of the renormalized gauge coupling. 
Its precise relation to the microscopic gauge couplings $g$ can typically be influenced by the presence of scalar fluctuations with masses of the order of the inverse lattice distance. 
Thus the relation \eqref{23} can be modified for finite $\lambda$, while the overall picture remains the same as long as these couplings are large enough.

\medskip\noindent
{\bf 6. Site scalars and interactions with link variables}

We finally turn to the remaining part of the link-scalar model which involves the site scalars $\chi$ and the interaction between $L$ and $\chi$.
We show in appendix D that the terms $-\mathcal{L}_p$ and $-\mathcal{L}_l$ in the scalar action \eqref{30C} are canceled by corresponding terms in $S_L[-\tilde L].$
The resulting action of the link scalar model reads
\ba\label{84A}
S[\rchi,L]&=&\sum_x\left(d\mu^2\tr M^2+\frac{dN\lambda}{2}\tr M^4\right)+\sum_{\rm plaquettes}\!\!\!\!{\cal A}_p\nn\\
&&+S_L[L]+S_{\rm int}[\rchi,L],
\ea
and we note the simple form of the part containing only the scalars $\chi$.

The interaction between scalars and link variables is given by
\ba\label{A4}
S_{{\rm int}}[\rchi,L]&=&S_L\big [L_{ij}\m -\tilde L_{ij}\m\big]-S_L\big[L_{ij}\m\big]\nn\\
&&-S_L\big[\tilde L_{ij}\m\big].
\ea
Details of the computation of $S[\rchi,L]$ and generalizations are given in appendix D. 
We emphasize that the link-scalar model with action \eqref{84A} is completely equivalent to the scalar lattice gauge theory with action \eqref{30C}.

For the link-scalar model the potential and derivative terms in the scalar part of the action \eqref{84A} are simple, without the competition of terms $\sim -{\cal L}_p$ and $-{\cal L}_l$. 
In particular, both terms in the potential for $M$ are positive, with minimum at $M=0$. 
In general, the coupling between links and scalars is rather complicated, however. 
The piece $S_{\rm int}$ involves terms with up to three powers of $L$ and up to six powers of $\chi$. 
The terms quadratic in $\chi$ obtain from a Taylor expansion of $S_L[L]$
\ba\label{FA}
S^{(2)}_{\rm int}&=&-\sum_{\rm links}\Big\{\frac{\partial S_L[L]}{\partial L_{ij}(x;\mu)}
\tilde L_{ij}(x;\mu)\nn\\
&&\hspace{1.2cm}+\frac{\partial S_L[L]}{\partial L^\dagger_{ij}(x;\mu)}
\tilde L^\dagger_{ij}(x;\mu)\Big\}\\
&=&-\sum_{x,\mu}\chi^{a^*}_j(x+e_\mu)
\frac{\partial S_L}{\partial L_{ij}(x;\mu)}
\chi^a_i(x)+c.c..\nn
\ea
In particular, this quadratic term vanishes whenever $L$ corresponds to an extremum of $S_L[L]$. 
Terms with four powers of $\chi$ are proportional to the second $L$-derivative of $S_L$, while six powers of $\chi$ multiply the third derivative. 
We discuss more aspects of the interaction term $S_{\rm int}$ in the next section. 
There we show that the role of $S_{\rm int}$ is actually suppressed for large $\bar\lambda$ such that the main features of the link-scalar model can be understood in this limit. 

We can modify the Hubbard-Stratonovich type transformation \eqref{A2} by including in $\bar S_L$ also a dependence on the sources. 
This changes the relation between the derivatives of $W$ and the expectation values and correlation functions of link bilinears $\tilde L$. 
We discuss this issue in appendix E. 
In particular, one can find a choice of $\bar S_L$ such that the expectation values of link bilinears $\kl \tilde L\kr_\chi$ 
find a simple expression in terms of the expectation values $\kl L\kr$ for the link variables in the link-scalar model. 
Here $\kl \tilde L\kr_\chi$ refers to the expectation value in scalar lattice gauge theory with action \eqref{30C}. 
The expectation of the link bilinear $\kl \tilde L\kr_{\chi L}$ in the link scalar model is different (cf. app. E). 
We also discuss in appendix E alternative choices for the Hubbard-Stratonovich transformation. 

In conclusion of this section we have reformulated the scalar lattice gauge theory of sect. \ref{Invariant action and} into an equivalent link-scalar model with a functional integral over link variables. 
The action \eqref{30C} of scalar lattice gauge theory is equivalent to the action \eqref{84A} of the link-scalar model.

\section{Phase diagram of scalar lattice gauge theory}
\label{Phase diagram of scalar}
Scalar lattice gauge theory with collective link variables describes gauge bosons, scalars in the singlet and adjoint representation (from $L$) as well as flavored scalars in the fundamental representation $\chi$. 
Depending on the choice of the parameters in the action $S_\chi$ one expects a rich phase diagram, with the gauge sector in the confined or Higgs-phase, or without any long range interactions at all. 
The Higgs phase describes loosely speaking spontaneous symmetry breaking by expectation values of $\chi$ or $A_S$. 
(Spontaneous symmetry breaking of a local gauge symmetry occurs only in a gauge fixed version, being absent in a gauge invariant formulation. 
The relation between the different pictures is well known.)

It is not our aim to explore this phase diagram systematically in this paper. 
We mainly want to argue that for a suitable choice of $S_\chi$ one can realize the standard confinement phase of QCD or the Higgs phase for the electroweak $SU(2)\times U(1)$ gauge theory. 
For this purpose we present a recipe how to construct an action $S_\chi$ for scalar lattice gauge theory that is equivalent to the standard lattice gauge theory formulated in terms of unitary link variables. 

We employ the formulation as a link-scalar model by taking in eq. \eqref{A2} the choice $\bar S_L=S_L$. 
We concentrate on the action eq. \eqref{84A}. 
The action for the equivalent scalar lattice gauge theory is given by eq. \eqref{30C}.
This model has two free parameters, $\bar\lambda$ and $\bar m^2$, or, equivalently, $\lambda$ and $\mu^2$. 
(The two parameter sets are related by eqs. \eqref{XAX}, \eqref{XAY}.)

The qualitative properties of our model will depend on the role of the interaction term $S_{\rm int}[\chi,L]=S_{\rm int}[\tilde L,L]$ which is defined in eq. \eqref{A4}. 
Its possible importance can be seen from the identities
\be\label{A18}
S_{\rm int}[\tilde L=0,L]=0~,~S_{\rm int}[\tilde L=L,L]=-2S_L[L].
\ee
We will argue that the influence of the interaction term on the dynamics of the links remains small if the functional integral is dominated by configurations with small $\chi$.  

Consider now the limit $\lambda\to \infty,\mu^2\to\infty$ with fixed ratio
\be\label{80A}
l^2_0=\frac{\mu^2}{N\lambda}.
\ee
We first sketch the overall situation for this limit: 
The minimum of the action \eqref{84A} occurs for $\chi=0,M=0,\tilde L=0,L=l_0$. 
The fluctuations of $M$ and $\tilde L$ are strongly suppressed by quadratic terms $\sim \lambda$. 
More precisely, we may define a ``meson potential'' in the action \eqref{84A} by 
\be\label{59A}
V_W = d\mu^2 \tr M^2 + \frac{dN\lambda}{2} \tr M^4.
\ee
It strongly favors a value of M close to zero.
Similarly, the ``link-potential'' in $S_L$ suppresses the fluctuations of $L-l_0$ except for the pseudo-unitary link variables. 
Setting in the interaction term $S_{\rm int}[\tilde L,L]$ the link bilinears $\tilde L$ to zero makes this term vanish. 
Due to the suppression of $\tilde L$ for large $\lambda$ the interaction term is expected to play only a minor role in this limit. 
For $V_W$ and ${\cal A}_p$ also close to their minimum, $V_W=0, \:{\cal A}_p=0$, we are left with linear lattice gauge theory with action $S_L$. 
For this setting we have already shown in sect. \ref{Gauge bosons as collective excitations} that the limit $\lambda\to\infty$ with fixed $l^2_0$ reduces to a standard lattice gauge theory. 

We next discuss the suppression of $S_{\rm int}$ in more detail. 
The potential term $V_W$ in eqs. \eqref{59A}, \eqref{84A} reads 
\be\label{80B}
V_W=\frac{dN\lambda}{2}(2l^2_0\,{\rm tr}M^2+{\rm tr}M^4),
\ee
with minimum at $M=0$. Also ${\cal A}_p$, given by eq. \eqref{M4}, has its minimum for $M=0$. 
However, the potential term \eqref{80B} will dominate over ${\cal A}_p$ for $\lambda\to\infty$. 
The diverging quadratic term $\sim \tr M^2$ in eq. \eqref{80B} will suppress strongly the fluctuations of $M$. 
On the other hand, the link action $S_L$ takes its minimum for a homogeneous unit matrix $L\m=l_0$. 
Again, the diverging mass terms \eqref{24} strongly suppress the fluctuations of $l-l_0$ and $A_S$. 
Thus the dominant fluctuations around the minimum of $S_L$ are the pseudo-unitary link variables with 
\ba\label{73A}
L=l_0U.
\ea
If the interaction term $S_{\rm int}$ is indeed subleading, 
one expects to recover standard lattice gauge theory with unitary link variables and $g^2$ or $\beta$ given by eq. \eqref{23}. 

For an estimate of the importance of the interaction term $S_{\rm int}$ we first note that it involves only terms with even powers of $\chi$. 
This implies that the total action $S[\chi,L]$ has an extremum for $\chi=0,L=l_0$. 
We next study the quadratic terms for excitations around this extremum. 
The fields $A_S,l-l_0$ and $\chi$ do not mix since they belong to different representations of the gauge group. 
The quadratic terms for $A_S$ and $l-l_0$ are not affected by the interaction term. 
They are large according to eq. \eqref{24}. 

In contrast, the quadratic term for $\chi$ could receive contributions from $S_{\rm int}$. 
By virtue of eq. \eqref{FA} they vanish, however, for the extremum of $S_L$ at $L\m=l_0$. 
For a positive quadratic term from an additional potential $\bar V(\chi)$ (cf. appendix E) one would find that the homogeneous configuration $\chi=0,L=l_0$ is indeed a local minimum of the link-scalar action. 
This extends to $\bar V(\chi)=0$. Now the dominant terms for small $\chi$ are quartic in $\chi$. 

For the term $\sim \chi^4$ the contribution from $S_{\rm int}$ corresponds to the terms $\sim \tilde L^2$ in the expansion of $S_L[L-\tilde L]-S_L[\tilde L]$. 
The leading terms arise from 
\be\label{73B}
\hat W_L=\frac{N\lambda}{2}\tr \big\{\big[(L-\tilde L)^\dagger(L-\tilde L)\big]^2\big\}.
\ee
With $L$ approximated by $l_0$ this yields for the term $\sim \chi^4$
\be\label{73C}
\hat W^{(4)}_L=\frac{N\lambda}{2}l^2_0\,\tr \big\{(\tilde L^\dagger+\tilde L)^2+2\tilde L^\dagger\tilde L\big\}.
\ee
This term is positive and its minimum occurs for $\tilde L=0$ or $\chi=0$. 
Since also the minimum of the potential term \eqref{80B} is found at $\chi=0$ we infer that the minimum of $S[\chi,L]$ indeed occurs for $\chi=0$. 

The term \eqref{73C}, together with the potential \eqref{80B}, strongly suppresses fluctuations of $\tilde L$ such that the role of the interaction term is indeed negligible. 
In the limit $\lambda\to \infty,l^2_0$ fixed, we can  approximate $S[\chi,L]$ by setting $\chi=0$. 
The link-scalar model \eqref{84A} reduces to linear lattice gauge theory. 
Following sect. \ref{Gauge bosons as collective excitations} it corresponds to standard lattice gauge theories with microscopic gauge coupling given by eq. \eqref{23}.

The minimum at $\chi_0 = 0$, $L=l_0$ is an important feature for the Hubbard-Stratonovich type transformation to be useful.
Keeping in the action $S[\chi, L]$ for the link-scalar model only the part $S_L[L-\tilde L]$ would imply an extremum at $\tilde L = L$.
It is the combination of this piece with the part $S^{(\chi)}$ in eq. \eqref{30C} which leads to the structure of the minimum for vanishing $\chi$ and nonzero $L$.
The form of $S^{(\chi)}$ is therefore important.
In particular, an expansion of $S_L[L-\tilde L]$ contains a quadratic term in $\tilde L$ in addition to the expression \eqref{73C},
\be\label{63A}
\Delta\hat W_L^{(4)} = -\mu^2\,\tr\tilde L^\dagger \tilde L = -N\lambda l_0^2 \operatorname{tr} \tilde L^\dagger \tilde L.
\ee
This cancels the last term in eq. \eqref{73C}, such that $\hat W_L^{(4)}+\Delta\hat W_L^{(4)}$ vanishes for antihermitean $\tilde L$.
However, the term $\hat W_L^{(4)}$ is canceled by the piece $-\mathcal{L}_l$ in $S^{(\chi)}$.
We may consider 
\ba\label{63B}
\bar W(\tilde l) &=& = \left. W_L (L-\tilde L) \right|_{L=l_0,\,\tilde L=\tilde l} \nn \\
&=& N^2 \lambda \left(-\frac12 l_0^4 + 2l_0^2\tilde l^2 - 2l_0\tilde l^3+\frac12 \tilde l^4 \right),
\ea
which has to generate minima at $\tilde l = 0$ and $\tilde l = 2l_0$, with a maximum at $\tilde l = l_0$.
The addition of $S^{(\chi)}$ preserves the minimum at $\tilde l = 0$, while the one at $\tilde l = 2l_0$ disappears.
(Note that for $N_f < N$ the choice $\tilde L = \tilde l$ would not be possible since $\det \tilde L = 0$.)

To summarize this discussion, we conclude that the limit 
\be\label{73Ca}
\bar\lambda\to\infty~,~\bar m^2=2d\bar\lambda l^2_0a^{-\frac{d+2}{2}}
\ee
of scalar lattice gauge theory with action \eqref{30C} describes standard lattice gauge theory. 
Only the precise relation between correlation functions of unitary link variables in standard lattice gauge theories and correlation functions for link bilinears in scalar lattice gauge theory is complicated, 
as we discuss in appendix E. 
One expects that the same universality class extends to finite large values of $\bar\lambda$. 
Other universality classes with additional light degrees of freedom may be realized for small $\bar\lambda$ or in the presence of additional terms in the action of scalar lattice gauge theory. 

While for large $\bar\lambda$ a guess of the universality class seems rather straightforward in the formulation with scalars and link variables the situation is less obvious 
in the equivalent formulation which only uses scalars and action $S_\chi$ as given by eq. \eqref{30C}. 
This formulation exhibits the complicated derivative interactions given by $\bar S_p$. 
It is not very clear a priori which is the dynamics induced by these derivative interactions which involve eight powers of $\chi$ at up to four different lattice sites. 
The reformulation with link variables is therefore very helpful for a clarification of the situation. 

We end this section by a brief discussion how the Higgs phase can be implemented within scalar lattice gauge theory. 
In order to achieve spontaneous symmetry breaking by a scalar in the fundamental representation of $SU(N)$ we add a potential $\bar V(\chi)$ to the action \eqref{30C} or \eqref{84A}. 
We take a term quadratic in $\chi$ or linear in $M$
\be\label{XH1}
\bar V_2=\tr \{M m^T\}=(\chi^a_i)^* m^{ab}\chi^b_i.
\ee
For a general ``mass matrix'' $m$ this breaks the flavor symmetry. 
We  may now choose $m$ such that after the inclusion of quantum fluctuations the corresponding renormalized mass matrix has one negative eigenvolume $-m^2_-$, while the other eigenvalues are large and positive. 
For the electroweak theory the eigenvector corresponding to the negative eigenvalue plays the role of the Higgs doublet. 
Its expectation value is set by $m^2_-/\lambda_-$, with $\lambda_-$ some appropriate quartic coupling. 
For a small scale of electroweak symmetry breaking (compared to $a^{-1}$) one has to choose parameters such that $m^2_-a^2$ is small. 
The renormalized quartic coupling $\lambda_-$ receives contributions from the terms \eqref{80B} and \eqref{73C}. 
For large $\lambda$ one expects large $\lambda_-$. 
Smaller couplings $\lambda_-$ can be realized by choosing small $\lambda$ or by adding additional terms in the form of $\bar V(\chi)$. We emphasize that the flavor symmetry plays no essential role in our arguments. There is no need 
that $\bar V(\chi)$ respects this symmetry, allowing for considerable freedom.

\section{Gauge invariant formulation}
\label{Gauge invariant formulation}

The action of scalar lattice gauge theory involves only the gauge invariant matrix $M$. 
One may therefore wish to reformulate the model in terms of a functional integral only involving the gauge invariant variables $M$. 
This gets rid of the gauge degrees of freedom altogether. 
While this reformulation is possible, it introduces a rather complicated non-polynomial potential for $M$. 
In particular, this potential diverges whenever $M$ cannot be represented as a bilinear in $\chi$ according to eq. \eqref{1A}., thus excluding effectively those values from the functional integral over $M$.

One can achieve such a formulation by inserting in the functional integral an integration over a product of $\delta$-functions 
\ba\label{B1}
&&\int {\cal D}M\prod_x\delta\big(M(x)-\chi^\dagger(x)\chi(x)\big)=1,\nn\\
&&\int{\cal D}M=\prod_x\int dM(x).
\ea
Here $\int dM$ denotes an integral over the $N_f^2$ independent components of the hermitean $N_f\times N_f$ matrix $M(x)$. 
Since the action only depends on $M(x)$ this yields
\ba\label{B2}
Z&=&\int{\cal D}\chi\exp \big\{-S_\chi\big[M(\chi)\big]\big\}\nn\\
&=&\int {\cal D}M\exp \big\{-S_\chi[M]-
\sum_xV_M\big(M(x)\big)\big\},
\ea
with 
\ba\label{B3}
V_M(M)=-\ln B(M)~,~B(M)=\int d\chi d\chi^\dagger \delta(M-\chi^\dagger\chi).\nn\\
\ea

A few properties of $B(M)$ can be found from simple arguments. 
The matrix $\chi^\dagger \chi$ obeys constraints - for example, the diagonal elements cannot be negative, or $\det \chi^\dagger\chi=0$ if $N_f>N$. 
If the integration variable $M$ does not obey the same constraints a solution $M=\chi^\dagger\chi$ does not exist and $B(M)=0$. 
Thus the potential $V(M)$ diverges if $M$ does not obey the constraints, restricting effectively the allowed values of $M$ to such matrices that can be represented as $\chi^\dagger \chi$. 

We next investigate matrices $M$ for which $B(M)\neq 0$, and begin with diagonal matrices. 
From a rescaling $\chi\to\alpha\chi,M\to\alpha^2M$ one infers $B(\alpha^2M)=\alpha^{2N_fN-2}B(M)$ and therefore $B\sim M^{N_fN-1}$. 
Since $B$ is positive one infers for $N_fN>1$ that $V_M$ diverges for $M\to 0$, $V_M(M\to 0)\to \infty$. 
Similarly, $V_M$ diverges logarithmically to negative values for large $M$. We note the special case $N_f=N=1$ with an abelian gauge symmetry. 
In this case $B=\pi$ is constant and $V_M$ can be omitted. 
As we have discussed in app. B this case is not a usual abelian gauge theory. 

Furthermore, $B$ is invariant under $SU(N_f)\times U(1)$ flavor transformations $M\to \hat U^\dagger M\hat U$. 
Indeed, such transformations of $M$ can be accompanied by a transformation $\chi\to\chi\hat U$, leaving $B$ invariant. 
In consequence, $B(M)$ can be written in terms of flavor invariants as ${\rm tr}M^2,{\rm tr}M^4$ etc.. 
This allows one to extract the full $M$-dependence of $B$ from its evaluation for diagonal matrices.

In principle, the expectation values of all observables $O[M]$ that can be expressed in terms of $M$ can be computed by inserting $O[M]$ in the functional integral \eqref{B2}. 
This includes generalized plaquette or loop observables which consist of an ordered matrix product of links along a closed loop. 
Such observables can be expressed by an ordered matrix product of factors $M$ at each point of the loop. 
Important properties of gauge theories can be investigated in a formulation employing only gauge singlets where the gauge transformations are no longer visible at all!

Besides the formulation of scalar lattice gauge theories in terms of complex scalar fields $\chi$, and the equivalent formulation with additional link variables $L$, 
one therefore could also employ a third equivalent ``meson formulation'' in terms of a functional integral involving only the gauge singlets $M$. 
No trace of the gauge transformations is directly visible anymore - the only memory is the additional potential $V_M$. 
Since the matrices $M$ transform non-trivially under the global flavor symmetry $SU(N_f)\times U(1)$ we may also call the functional integral in terms of $M$ the ``flavor formulation''. 
In a wider sense, this corresponds to a formulation of gauge theories in terms of generalized ``mesons''. 

For $N_fN>1$ the presence of the potential $V_M$ in the microscopic action for $M$
\ba\label{B4}
S_M[M]&=&S_\chi[M]+S_V[M],\nn\\
S_V[M]&=&\sum_xV_M\big(M(x)\big)
\ea
has an important consequence. 
The combined potential $\hat V(M)=2d\mu^2\tr M^2+V_M(M)$ has its minimum for $M_0\neq 0$.

Despite the conceptual simplicity of a formulation in terms of gauge invariant fields $M(x)$ the specific form of the potential $V_M(M)$, 
which encodes the constraints for $M$, makes the gauge invariant formulation difficult to handle in practice for many purposes. 
For scalar lattice gauge theory with unconstrained fields $\chi(x)$ the polynomial form of the action is simpler. 
In particular, the connection to link bilinears is much more direct if one uses the scalar fields $\chi(x)$. 
From the point of view of the meson formulation the constraints on $M(x)$ are precisely what is needed in order to permit a reformulation in terms of $\chi(x)$. 

The meson formulation highlights that even for scalar lattice gauge theory formulated in terms of $\chi$ 
the dominant configurations for the functional integral are not around $\chi=0$, but rather around the minimum of $\hat V$. 
This is due to the functional measure which suppresses the weight of fluctuations with very small $\chi$. 
Still, for $\lambda\to\infty,\,\mu^2\to\infty$ the location of the minimum of $\hat V$ approaches zero.

\section{Conclusions}
\label{Conclusions}

We have presented a formulation of scalar lattice gauge theory. 
It is based on scalar fields, located on lattice sites, rather than on unitary (or orthogonal) link variables as in standard lattice gauge theories. 
The action and functional measure are invariant under local gauge transformations. 
We have found a limit for the parameters of scalar lattice gauge theory for which the long-distance behavior is the same as for standard lattice gauge theories. 
In particular, one expects the usual confinement property for pure non-abelian gauge theories. 
This shows that gauge fields can be obtained as collective excitations or composites of ``fundamental'' scalar fields. 

Varying the parameters of scalar lattice gauge theory one expects a rich phase diagram, with confinement as well as Higgs phases. 
In particular, the gauge group $SU(3)\times SU(2)\times U(1)$ of the standard model, with $SU(3)$ in the confinement and $SU(2)\times U(1)$ in the Higgs phase, can be represented as a scalar lattice gauge theory. 
The ``fundamental'' degrees of freedom are scalars in the fundamental representation of the gauge group. 
Scalar lattice gauge theory is accessible to numerical simulations by standard Monte-Carlo methods. 
An exploration of the phase diagram may shed new light on different universality classes for gauge theories and their mutual connection. 

As an intermediate step we have encountered linear lattice gauge theory. 
This type of model uses unconstrained linear link variables instead of the usual unitary or orthogonal link variables. 
Linear lattice gauge theory is interesting in its own right. 
In particular, the inverse gauge coupling is proportional to the expectation value $l^2_0$ of the squared linear link variable. 
It seems possible that confinement in the non-linear version of a gauge theory can be associated to the ``symmetric regime'' in the linear version of a gauge theory where $l_0$ vanishes. 
For a vanishing expectation value of the link variable the gluons are not propagating degrees of freedom. 

An extension to ``fundamental'' fermions instead of scalars can be made along the lines discussed in this paper. 
In such a ``fermion lattice gauge theory'' Grassmann variables $\psi^a_{i,\alpha}(x)$ and $\bar\psi^a_{i,\alpha}(x)$ replace the scalar field $\chi^a_i(x)$ and $(\chi^a_i(x))^*$. 
The index $\alpha$ is an additional Lorentz index, e.g. $\alpha=1\dots 4$ for Dirac spinors in $d=4$. 
The functional integral \eqref{1A} or \eqref{30} is now a Grassmann functional integral. 
The matrices $M^{ab}(x)$ are replaced by the gauge singlets $M^{ab}_{\alpha\beta}(x)=\bar\psi^a_{i,\alpha}(x)\psi^b_{i,\beta}(x)$. 
In particular, the combinations $M^{ab}_{\alpha\beta}\delta^{\alpha\beta}$ and $M^{ab}_{\alpha\beta}\bar\gamma^{\alpha\beta}$ (with $\bar\gamma$ the generalization of $\gamma^5$) 
transform as scalars or pseudoscalars and can be associated with mesons. 
Bilinear link variables $\tilde L_{ij}$ are constructed similar to eq. \eqref{7} as Lorentz and flavor singlets. 
One may construct a gauge invariant action as a polynomial of $M$ which is now an element of the Grassmann algebra constructed from $\psi$ and $\bar\psi$. 
One expects again to find a region in parameter space where the universality class is the same as a gauge theory coupled to fermions. 
In this way one may realize QCD with quarks in a formulation which only involves fermionic fields. 
Even further, the standard model of particle physics can be formulated by using only fermions as fundamental degrees of freedom.

\medskip\noindent
{\bf Acknowledgment}

\noindent
The author would like to thank D.~Horn, M.~Luscher and U.~Wiese for communicating results of their earlier work.

\section*{APPENDIX A: ACTIONS WITH LOCAL GAUGE INVARIANCE}
\renewcommand{\theequation}{A.\arabic{equation}}
\setcounter{equation}{0}

\noindent
{\bf 1. Local invariants}

In this appendix we display details and extensions for the action of scalar lattice gauge theory. 
The action $S[\chi]$ is invariant under local gauge transformations if it only involves local invariants, as
\ba\label{1}
M^{ab}(x)&=&\big(\chi^a_i(x)\big)^*\chi^b_i(x)~,~I(x)=\big(\chi^a_i(x)\big)^*\chi^a_i(x),\nn\\
R^{abcd}(x)&=&\big(\chi^a_i(x)\big)^*(\lambda_z)_{ij}\chi^b_j(x)\big(\chi^c_k(x)\big)^*(\lambda_z)_{kl}\chi^d_l(x).\nn\\
\ea
Here $\lambda_z$ are the generators of $SU(N)$ and summation over repeated indices is implied. 
The precise local gauge symmetry depends on which local invariants appear in the action. 
(Note $I=$tr$M$ if $M^{ab}$ is interpreted as an $N_f\times N_f$ matrix.) If $S$ involves only $M^{ab}$ and $R^{abcd}$ the local gauge symmetry is $SU(N)\times U(1)$. 
We may also consider real scalar fields - $R^{abcd}$ is not defined in this case. If $S$ can be written on the level of the real bilinears $M^{ab}(x)$ the local gauge symmetry is $SO(N)$. 

Invariants that are singlets with respect to $SU(N)$, but charged with respect to the $U(1)$-subgroup, involve $N$ in powers of $\chi$ (for $N_f\geq N)$
\be\label{2AB}
D^{a_1a_2\dots a_N}(x)=\epsilon^{i_1i_2\dots i_N}\chi^{a_1}_{i_1}(x)\chi^{a_2}_{i_2}(x)\dots \chi^{a_N}_{i_N}(x). 
\ee
Models with action formulated in terms of $M,R$ and $D$ are invariant under local $SU(N)$-gauge transformations. 
Furthermore, one has a local $U(1)$ symmetry if each term in the action has an equal number of factors $D$ and $D^*$ at every point $x$. 
In this case only the relative values of the gauge couplings of $SU(N)$ and $U(1)$ will be modified by the presence of terms involving $D$. 

Terms with derivatives of $D$ may not involve an equal number of factors of $D$ and $D^*$ at each point $x$. 
Then the local symmetry is reduced to $SU(N)$. A global $U(1)$ symmetry is still preserved if globally the number of factors $D$ and $D^*$ is equal for each terms. 
Breaking the local $U(1)$ symmetry by terms involving $D$ will be a way to construct pure QCD based on the gauge group $SU(3)$. 
In addition to the local gauge symmetry one may have additional global symmetries acting on the flavor indices. 

We concentrate on models with local $SU(N)\times U(1)$ gauge symmetry. 
Let us consider different types of terms that may appear in the action. 
A potential term is given by 
\be\label{2}
S_V=\sum_xV\big(M^{ab}(x),R^{abcd}(x), D^*(x)D(x)\big).
\ee
In the following we will not use $R$, nor other possible higher order invariants as $D^*D$. 
They can be added if needed. If the action contains only $S_V$ the model is a simple ultra-local theory where different lattice points are independent of each other. 

We next introduce
\ba\label{3}
K^{ab}_\mu(x)&=&M^{ab}(x+e_\mu)-M^{ab}(x)\\
&=&\big(\chi^a_i(x+e_\mu)\big)^*\chi^b_i(x+e_\mu)-\big(\chi^a_i(x)\big)^*\chi^b_i(x),\nn
\ea
with $e_\mu$ a unit vector in one of the $d$ lattice directions. 
A term 
\be\label{4}
S_K=\sum_x\big(K^{ab}_\mu(x)\big)^*K^{ab}_\mu(x)
\ee
links different lattice sites such that the partition function is no longer a trivial product of individual site contributions. 
Besides local gauge invariance this term is invariant under global (not local!) $SU(N_f)$ flavor transformations, as well as under lattice rotations by $\pi/2$. 
(We discuss here euclidean lattice gauge theories, with a possible generalization to Minkowski signature or, alternatively, analytic continuation of the correlation functions to Minkowski space.)

Using the lattice derivatives \eqref{5} one sees that $S_K$ amounts to a kinetic term for the invariant $M^{ab}$
\be\label{6}
S_K=a^2\sum_x\partial_\mu\big(M^{ab}(x)\big)^*\partial_\mu M^{ab}(x).
\ee
This can be generalized to other terms involving derivatives of the local invariants $M$ or $R$. 
An example is the action \eqref{30C}.

\medskip\noindent
{\bf 2. Link bilinears}

We can express $S_K$ partly in terms of the link bilinears \eqref{7}, using 
\ba\label{8}
&&\big(K^{ab}_\mu(x)\big)^*K^{ab}_\mu(x)=
\big(M^{ab}(x+e_\mu)\big)^*M^{ab}(x+e_\mu)\nn\\
&&+\big(M^{ab}(x)\big)^*M^{ab}(x)
-2\tilde L_{ij}(x;\mu)\tilde L^*_{ij}(x_i;\mu),
\ea
(no summation over $\mu$ here). 
This yields $S_K$ as a sum over links denoted by $(x;\mu)$, plus a potential term, 
\ba\label{9AA}
S_K=2d\sum_x\big(M^{ba}(x)\big)^\dagger M^{ab}(x)\nn\\
-2\sum_{{\rm links}}\tilde L^\dagger_{ji}\m\tilde L_{ij}\m.
\ea
Here the sum over links amounts to $\sum_x\sum_\mu$. 

We have already introduced in sect. \ref{Invariant action and} the plaquette invariants $\tilde P\m$, cf. eq. \eqref{11}. 
They correspond to a product of links around a plaquette. A further type of invariants,
\ba\label{14}
&&\tilde Q(x;\mu,\nu)=\tilde Q(x;\nu,\mu)=\tilde Q^*(x;\mu,\nu)\nn\\
&&\quad ={\rm tr}\big\{\tilde L(x;\mu)\tilde L^\dagger(x;\mu)\tilde L(x;\nu)\tilde L^\dagger(x;\nu)\big\},
\ea
corresponds to a sequence of links from $x$ to $x+e_\mu$ and back, and then to $x+e_\nu$ and back. The combination
\ba\label{15}
&&{\cal L}_p(x;\mu,\nu)=-\frac12\big[\tilde P(x;\mu,\nu)+\tilde P(x;\nu,\mu)\big]\nn\\
&&\quad +\frac14\big[\tilde Q(x;\mu,\nu)+\tilde Q(x+e_\mu;\nu,-\mu)\\
&&\quad +\tilde Q(x+e_\mu+e_\nu;-\mu,-\nu)+\tilde Q(x+e_\nu;-\nu,\mu)\big]\nn
\ea
can be viewed as a type of locally gauge invariant kinetic term for the collective link variables. 
It can be written in the form \eqref{23FA}, such that the positivity ${\cal L}_p\geq 0$ is manifest.

We can express $\tilde Q$ in terms of $M$ as 
\be\label{16A}
\tilde Q(x;\mu,\nu)={\rm tr}\big\{M(x)M(x+e_\mu)M(x)M(x+e_\nu)\big\}.
\ee
In terms of the scalar invariants $M$ one finds
\ba\label{16B}
{\cal L}_p&=&\frac14{\rm tr}\big\{M(x+e_\nu)\big[M(x+e_\mu+e_\nu)-M(x)\big]\nn\\
&&\times M(x+e_\mu) \big[M(x+e_\mu+e_\nu)-M(x)\big]\nn\\
&&+M(x)\big[M(x+e_\mu)-M(x+e_\nu)\big]M(x+e_\mu+e_\nu)\nn\\
&&\times\big[M(x+e_\mu)-M(x+e_\nu)\big]\big\}.
\ea
This coincides with eq. \eqref{M2}. 

In this paper we investigate models which involve ${\cal L}_p$ with a negative sign. 
In order to have a plaquette action that is bounded from below we have to combine $-{\cal L}_p$ with some other piece, $\bar\s_p={\cal A}_p-{\cal L}_p\geq 0$.  
With hermitean matrices $M$ and 
\ba\label{B3c}
A_\pm&=&M(x+e_\mu)\pm M(x+e_\nu),\nn\\
B_\pm&=&M(x+e_\mu+e_\nu)\pm M(x), 
\ea
one has
\ba\label{23AA}
{\cal L}_p&=&\frac{1}{16}{\rm tr}
\{A_+B_-A_+B_-+B_+A_-B_+A_-\nn\\
&&-2A_-B_-A_-B_-\}.
\ea
We define 
\be\label{23AB}
{\cal A}_p=\frac{1}{16}{\rm tr}\{A^2_+B^2_-+B^2_+A^2_-+2A^2_-B^2_-\}\geq 0.
\ee
For hermitean matrices $K,L$ one has the identities
\ba\label{24AC}
&&{\rm tr}\{K^2L^2-KLKL\}=\frac12{\rm tr}\{[K,L]^\dagger [K,L]\},\nn\\
&&{\rm tr}\{K^2L^2+KLKL\}=\frac12\tr \big\{\{K,L\}^\dagger \{K,L\}\big\},\nn\\
&&{\rm tr}\{K^2L^2\}=\frac14 {\rm tr}\big\{\{K,L\}^\dagger \{K,L\}\big\}\nn\\
&&\hspace{2.0cm}+\frac14{\rm tr}\big\{[K,L]^\dagger [K,L]\big\},
\ea
such that all three  combinations are positive semidefinite. 
This shows that the combination 
\ba\label{25AC}
\bar{\cal S}_p&=&{\cal A}_p-{\cal L}_p=
\frac{1}{16}{\rm tr}\{A^2_+B^2_--A_+B_-A_+B_-\nn\\
&&+B^2_+A^2_--B_+A_-B_+A_-\nn\\
&&+2A^2_-B^2_- +2A_-B_-A_-B_-\}\geq 0
\ea
is positive semidefinite. It coincides with eq. \eqref{M1}. 

\medskip\noindent
{\bf 3. Continuum limit}

One may express ${\cal L}_p$ and ${\cal A}_p$ in terms of lattice derivatives of $M$. For this purpose we note that $A_-$ and $B_-$ are derivatives,
\ba\label{16C}
&&M(x+e_\mu)-M(x+e_\nu)=a\big\{\partial_\mu M(x)-\partial_\nu M(x)\big\},\nn\\
&&M(x+e_\mu +e_\nu)-M(x)=\frac a2\big\{\partial_\mu\big[M(x+e_\nu)+M(x)\big]\nn\\
&&\qquad +\partial_\nu\big[M(x+e_\mu)+M(x)\big]\big\}.
\ea
We may also employ
\ba\label{16D}
&&\partial_\mu\partial_\nu M(x)=\partial_\nu\partial_\mu M(x)=\frac1a\partial_\nu\big[M(x+e_\mu)-M(x)\big]\nn\\
&&=\frac{1}{a^2}\big[M(x+e_\mu+e_\nu)-M(x+e_\mu)\nn\\
&&\quad-M(x+e_\nu)+M(x)\big]
\ea
in order to express ${\cal L}_p$ in terms of $M(x),\partial_\mu M(x),\partial_\nu M(x)$ and $\partial_\mu\partial_\nu M(x)$. 

The leading term in the continuum limit reads
\ba\label{16E}
{\cal L}_p(x;\mu,\nu)&=&\frac{a^2}{2}{\rm tr}\big\{M\partial_\mu MM\partial_\mu M+M\partial_\nu MM\partial_\nu M\big\}\nn\\
&&+0(a^4).
\ea
With $a^d\sum_x=\int d^dx=\int_x$ and an appropriate rescaling of $M,M\to a^{(d-2)/4}M$ this results in 
\be\label{23AAb}
S_p=\int_x\frac{d-1}{2}{\rm tr}\{M\partial_\mu MM\partial_\mu M\}.
\ee
For $d=4$ we recognize the last term in eq. \eqref{AA} as $-S_p$. Similarly, one finds in leading order 
\ba\label{30A}
{\cal A}_p(x;\mu,\nu)&=&\frac{a^2}{2}{\rm tr}\big \{M^2(\partial_\mu M\partial_\mu M+\partial_\nu M \partial_\nu M)\big\}\nn\\
&&+O(a^4).
\ea
This can be combined with the continuum limit of the link action 
\ba\label{A.23}
\bar\s_l&=&a^2\left(\bar\lambda-\frac{d-1}{2}\right) \tr \big\{M^2(\partial_\mu M)^2\big\}\nn\\
&&+\frac{\bar m^2}{d}a^{\frac{d+2}{2}}\tr M^2.
\ea
For $d=4$ this yields the naive continuum action \eqref{AA}. 
We emphasize, however, that the detailed lattice action is necessary for an understanding of our model. 
For example the crucial importance of the plaquette invariant $\tilde P$ is not visible in the naive continuum formulation. 

Finally, we observe that the rescaled continuum field $M_c(x)=M(x)a^{(2-d)/4}$ has dimension mass$^{(d-2)/4}$. 
For $d=4$ it scales $\sim \sqrt{{\rm mass}}$, different from the more familiar scaling of scalars $\sim$ mass. 
This is due to the absence of a quadratic kinetic term, while derivative terms involve four powers of $M$. 
Correspondingly, $\bar m^2$ has dimension mass$^3$ and $\bar\lambda$ is dimensionless. 
The rescaled continuum scalar field $\chi$ scales then with (mass)$^{1/4}$.

\medskip\noindent
{\bf 4. Extended setting}

The definition \eqref{1Aa} implies for the meson bilinears $M^{ab}(x)$ constraints as $M^\dagger=M$ or $M^{aa}\geq 0$. 
As a consequence, the minimum of the action \eqref{30C} occurs for $M(x)=0$. 
We can generalize our setting of scalar lattice gauge theory such that the meson fields are unconstrained and a minimum of the action can be realized also for $M(x)\neq 0$. 
For this purpose one may consider an extended set of fields $\chi^a_{\alpha i}(x)$, with $\alpha=1\dots S$ denoting an additional index for the $S$ ``species'' of scalars. 
We define the meson bilinears as
\be\label{U1}
M^{ab}(x)=\bar f^\alpha f^\beta \big(\chi^a_{\alpha i}(x)\big)^*\chi^b_{\beta i}(x).
\ee
An example is $S=2,\alpha=(+,-),\bar f_\alpha=(1,1),f_\alpha =(1,-1)$. The meson bilinears
\be\label{U2}
M^{ab}(x)=\big(\chi^a_{+i}(x)+\chi^a_{-i}(x)\big)^*\big(\chi^b_{+i}(x)-\chi^b_{-i}(x)\big)
\ee
are no longer hermitean and $M^{aa}$ can take complex values. (For $\chi^a_{+i}=0$ one has real negative $M^{aa}$.) 
As before, the meson bilinears are invariant under local gauge transformations. 
Link bilinears are now defined as
\be\label{U3}
\tilde L_{ij}\m=f^\alpha\bar f^\beta \chi^a_{\alpha i}(x)\big(\chi^a_{\beta j}(x+e_\mu)\big)^*.
\ee
They have the standard transformation property \eqref{10} with respect to local gauge transformations. 
In the present paper we will not pursue this generalization and rather stick to $S=1$ with $M$ defined by eq. \eqref{1Aa}.

\section*{APPENDIX B: PHASE FLUCTUATIONS AND CRITICAL FLAVOR NUMBER}
\renewcommand{\theequation}{B.\arabic{equation}}
\setcounter{equation}{0}

The dynamics of lattice gauge theories is closely related to phase fluctuations. 
Indeed, the standard lattice gauge theories are formulated in terms of link variables that are unitary matrices. 
These unitary matrices can be considered as generalized phases. In scalar lattice gauge theories the issue of phases depends on the number of flavors $N_f$. 
This will impose restrictions on $N_f$ if we want to obtain the same universality class as standard lattice gauge theories.

For our choice $N_f=2dN$ arbitrary configurations of the link bilinears $\tilde L\m$ can be realized rather easily for suitable scalar configurations $\chi(x)$.
Suppose that all $\chi (x)$ except for one particular point $x_0$ are given.
This fixes all link bilinears except for the $2d$ links that start or end at $x_0$.
We now want to choose the scalar field value $\chi(x_0)$ such that the $2d$ link bilinears $\tilde L(x_0;\mu)$, $\tilde L (x_0+e_\mu;-\mu)$ take prescribed values.
Since $\tilde L$ are complex $N \times N$ matrices this yields a system of $2dN^2$ linear complex equations for the $N_f N =2dN^2$ complex components of $\chi(x_0)$.
They can be solved for $\chi(x_0)$ as a function of $\chi(x_0\pm e_\mu)$, $\tilde L\m$, $\tilde L(x+e_\mu;-\mu)$, except for particular ``singular'' cases.

The singular cases depend on the values of the scalar field at neighbouring sites $x_0\pm e_\mu$.
For a singular case one may choose a different configuration for the scalars $\chi(x\neq x_0)$ which lead to the same link bilinears for all links not starting or ending at $x_0$.
Indeed, for one particular neighbor of $x_0$ at $x_1$ the value $\chi(x_1)$ has only to obey $(2d-1)N^2$ equations that determine it for given $\chi(x)$ at all points except $x_0$ and $x_1$
in terms of the links starting or ending at $x_1$, except for the one link joining $x_0$ and $x_1$ which is still free.
The corresponding $(2d-1)N^2$ equations for $2dN^2$ unknowns $\chi(x_1)$ have, in general, more than one solution.
This construction can be continued to more extended regions by induction.
The manifold of configurations $\chi(x)$ that realize a given configuration of link bilinears $\tilde L\m$ is further extended for $N_f>2dN$.
If arbitrary $\tilde L$ can be realized one can also realize arbitrary $\tilde U$ in eq. \eqref{15A1}.
The phase information of standard lattice gauge theory is contained in values of the scalars $\chi$.
For a suitable action one expects that scalar lattice gauge theory belongs to the same universality class as standard lattice gauge theories.

For smaller number of flavors $N_f$ the phase information contained in unitary link variables is no longer necessarily available. 
As a first example we consider the case $N=1$ with abelian $U(1)$-gauge symmetry. 
For $N_f=1$ the scalar $\chi$ is a single complex field that we may write as $r(x) \exp [i\alpha(x)]$, with real $r\geq 0$. 
The definitions \eqref{1Aa}, \eqref{7} imply
\ba\label{P1}
M(x)&=&r^2(x),\nn\\
\tilde L\m&=&r(x)r(x+e_\mu)e^{i\big(\alpha(x)-\alpha(x+e_\mu)\big)}.
\ea
It is obvious that the phases $\alpha(x)$ do not appear in the action - they are pure gauge degrees of freedom. 
We can define the phase of the link bilinear $\tilde L\m$ by $\beta\m=\alpha(x)-\alpha(x+e_\mu)$. 
The sum of phases $\beta$ for a product of link bilinears around a plaquette is constrained to vanish. 
Such a model is not expected to share similar properties as a standard abelian gauge theory.

The situation is different for $N_f=2$. We have now two complex scalar fields $\chi^a(x)$, and obtain with the two phases $\alpha_{1,2}(x)$
\ba\label{P2}
\chi^a(x)&=&r_a(x)\exp [i\alpha_a(x)],\nn\\
M^{ab}(x)&=&r_a(x)r_b(x)\exp \big[i\big(\alpha_b(x)-\alpha_a(x)\big)\big],\\
\tilde L\m&=&r_1(x)r_1(x+e_\mu)\exp \big[i\big(\alpha_1(x)-\alpha_1(x+e_\mu)\big)\big]\nn\\
&&+r_2(x)r_2(x+e_\mu)\exp \big[i\big(\alpha_2(x)-\alpha_2(x+e_\mu)\big)\big].\nn
\ea
A non-trivial phase $\alpha_2-\alpha_1$  is present in $M$. For the phase $\beta\m$ of the link variable one has 
\ba\label{P3}
&&\tan\beta\m\\
&&=\frac{\Sigma_ar_a(x)r_a(x+e_\mu)\sin [\alpha_a(x)-\alpha_a(x+e_\mu)]}
{\Sigma_ar_a(x)r_a(x+e_\mu)\cos [\alpha_a(x)-\alpha_a(x+e_\mu)]}.\nn
\ea
There is no constraint anymore that the sum of $\beta$ around a plaquette has to vanish. 
If the phase fluctuations play a decisive role such a model may be expected to belong to the same universality class as a standard abelian lattice gauge theory. 

These findings can be generalized to arbitrary $N$. 
For any given $N$ we expect that there is a critical flavor number $N_{f,c}(N)$ 
such that for $N_f\leq N_{f,c}$ the phase fluctuations are too much constrained such that scalar lattice gauge theory cannot belong to the standard universality class of gauge theories. 
We have seen already that this is the case for $N_f=1$ for arbitrary $N$, such that $N_{f,c}(N)\geq 1$. 
In the opposite, for $N_f>N_{f,c}(N)$ there exists a region in the parameter space of scalar lattice gauge theory for which the universality class of a standard gauge theory is realized. 

An estimate of $N_{f,c}(N)$ is a difficult task. 
In order to get some intuition, we present here some simple counts of degrees of freedom. 
First, we observe that for a periodic lattice with ${\cal N}$ sites the total number of real degrees of freedom contained in $\chi$ is $2NN_f{\cal N}$. 
On the other hand, for unconstrained link variables the total number of real degrees of freedom is $2N^2d{\cal N}$. 
For $N_f<dN$ the link bilinears $\tilde L\m$ defined by eq. \eqref{13B} are not all independent but rather have to obey constraints. 
One may speculate that for $N_f\geq dN$ (and $d>1$) the phase fluctuations are always sufficiently unconstrained in order to admit the universality class of standard gauge theories. 
This would imply $N_{f,c}(N)<dN$. 

For $N_f=dN$ and generic link configurations $\tilde L$ it will be possible to find scalar configurations $\chi$ obeying eq. \eqref{7}.
The system of quadratic equations involves, however, the fields $\chi(x)$ for all $x$ simultaneously.
For this reason our example $N_f = 2dN$ doubles the degrees of freedom in $\chi$ such that only local linear equations need to be solved, as discussed at the beginning of this appendix.

A second counting concerns the degrees of freedom appearing in a simple plaquette term ${\cal L}_p$. 
There are four sites and therefore $8NN_f$ real degrees of freedom in the variables $\chi(x)$. 
Due to the local gauge symmetry not all of them appear in ${\cal L}_p$. 
An upper bound for the number of independent degrees of freedom appearing in ${\cal L}_p$ is $4N(2N_f-N)$. 
(The number of real degrees of freedom in unconstrained hermitean fields $M(x)$ is $4N_f^2$. 
For $N_f>N$ the fields $M(x)$ obey constraints beyond hermiticity.) 
For unconstrained link variables we can use the gauge transformations in order to bring three link variables in the plaquette to a hermitean form, accounting for $3N^2$ degrees of freedom. 
This is not possible for the fourth link variable which retains the $2N^2$ degrees of freedom of an arbitrary complex matrix. 
The total number of link degrees of freedom appearing in ${\cal L}_p$ amounts therefore to $5N^2$. 
For $5N>8N_f-4N$ the links within each plaquette must obey local constraints. 
If these local constraints are strong enough to forbid the realization of the standard universality class one would infer $N_{f,c}(N)\geq (9/8)N$. 

An interesting case is $N_f=N$ for which $\chi,M$ and $\tilde L$ are all $N\times N$ matrices. 
We can write an arbitrary complex matrix $\chi$ in the form 
\be\label{P4}
\chi(x)=V(x)D(x)\tilde V^\dagger (x)
\ee
with real diagonal matrix $D(x)$ and unitary matrices $V(x)$ and $\tilde V(x)$. This implies 
\ba\label{P5}
M(x)&=&\tilde V(x)D^2(x)\tilde V^\dagger (x),\\
\tilde L\m&=&V(x)D(x)\tilde V^\dagger (x) \tilde V(x+e_\mu)D(x+e_\mu)V^\dagger(x+e_\mu).\nn
\ea
We can associate $V(x)$ with the gauge degrees of freedom that do not appear in the action. 
(One may set $V(x)=1.)$ The matrix $M(x)$ has positive eigenvalues given by $D^2(x)$. For $D^2(x)$ proportional to the unit matrix, $D^2(x)=m(x)$, one finds $M(x)=m(x)$. 
In this case the phases contained in $\tilde V(x)$ do not appear in the action. For different eigenvalues of $M(x)$, however, the action will depend on the phases $\tilde V(x)$. 
We emphasize in this context that for unitary $\chi$ the plaquette invariant \eqref{11} is a constant. A unitary link bilinear $\tilde L$ does not imply that $\chi$ is unitary. Nevertheless, one finds for generic $\chi$ that $M(x+\mu)M(x)=1$ such that $\tilde P$ is constant. Thus for $N_f=N$ and unitary links bilinears the plaquette invariant contains no phase information. This makes it unlikely that $N_f=N$ belongs to the standard universality class. Since the link bilinears need not to be unitary it is, however, not entirely clear if the standard universality class can be realized for $N_f=N$ or not. 

Finally we recall that for $N_f<N$ the determinant of the link bilinear vanishes, $\det\tilde L=0$. 
It seems likely that a constraint $\det\tilde L=0$ is not compatible with a usual gauge theory, implying $N_{f,c}(N)\geq N$. 
We will implicitly assume in this work that $N_f\geq N$.
While we concentrate in this paper on $N_f=2dN$, it remains an interesting question which universality classes can be realized for an ``intermediate flavor number'' $N\leq N_f < 2dN$.

\section*{APPENDIX C: FIELDS IN SCALAR LATTICE GAUGE THEORY}
\renewcommand{\theequation}{C.\arabic{equation}}
\setcounter{equation}{0}

In this appendix we discuss in more detail the degrees of freedom that are contained in the link bilinear $\tilde L$. 
This permits a formulation of a continuum limit of scalar lattice gauge theory in terms of collective gauge bosons and scalars. 
It also opens the way of constructing the continuum limit of the linear lattice gauge theory discussed in sec. \ref{Gauge bosons as collective excitations}.

\medskip\noindent
{\bf 1. Collective gauge bosons}

Collective link variables $\tilde L$ are defined as bilinears of scalar fields in eq. \eqref{7}. 
We can also introduce collective gauge fields $\tilde A_\mu(x)$ as suitable non-linear expressions of the scalars $\chi(x)$. For this purpose we write 
\be\label{80F1}
\tilde L\m=\tilde S(x)+ia\tilde B_\mu(x)
\ee
with (cf. eq. \eqref{AAB})
\be\label{80F2}
\tilde S_{ij}(x)=\chi^a_i(x)\chi^{a*}_j(x)=\big(\tilde S^\dagger(x)\big)_{ij}
\ee
and 
\ba\label{80F3}
(\tilde B_\mu)_{ij}(x)&=&-i\chi^a_i(x)\partial_\mu\chi^{a*}_j(x)\\
&=&-\frac ia\chi^a_i(x)
\big(\chi^{a*}_j(x+e_\mu)-\chi^{a*}_j(x)\big).\nn
\ea
In a matrix notation the bilinear continuum fields read 
\be\label{98A}
\tilde S=\chi\chi^\dagger~,~\tilde B_\mu=-i\chi\partial_\mu\chi^\dagger.
\ee
Thus $\tilde S$ are scalars and $\tilde B_\mu$ vectors. 
With respect to infinitesimal gauge transformations the continuum fields transform linearly
\be\label{80F4}
\delta\tilde S=i[\alpha,\tilde S]~,~\delta\tilde B_\mu=i[\alpha,\tilde B_\mu]-\tilde S\partial_\mu\alpha.
\ee
We note the unusual form of the term $\sim \partial_\mu\alpha$ which is multiplied by $\tilde S$. Therefore $\tilde B_\mu$ does not transform as a connection or gauge field.

The continuum collective gauge fields are defined by eq. \eqref{AAC},
\be\label{80F5}
\tilde A_\mu=\frac12(\tilde S^{-1}\tilde B_\mu+\tilde B^\dagger_\mu\tilde S^{-1}),
\ee
with standard inhomogeneous transformation property, cf. eq. \eqref{28F2},
\be\label{80F6}
\delta\tilde A_\mu=i[\alpha.\tilde A_\mu]-\partial_\mu\alpha.
\ee
We observe that $\tilde A_\mu$ exists only for regular $\tilde S$. (This requires $N_f\geq N$.)
As it should be the gauge fields are vectors.

In the continuum limit the bilinear $\tilde B_\mu$ contains also a second vector, associated to the antihermitean part of $\tilde S^{-1}\tilde B_\mu$, namely
\ba\label{80F10}
\tilde C_\mu&=&-\frac i2(\tilde S^{-1}\tilde B_\mu-\tilde B^\dagger_\mu\tilde S^{-1}),\nn\\
\tilde B_\mu&=&\tilde S(\tilde A_\mu+i\tilde C_\mu).
\ea
With respect to local gauge transformations $\tilde C_\mu$ transforms homogeneously
\be\label{80F11}
\delta \tilde C_\mu=i[\alpha,\tilde C_\mu].
\ee
(Recall that $\tilde S,\tilde A$ and $\tilde C_\mu$ are all hermitean matrices.)  
Using eqs. \eqref{80F1}, \eqref{80F10} the continuum expression of the collective link variables in terms of $\tilde S,\tilde A_\mu$ and $\tilde C_\mu$ reads in first order in a 
\be\label{80F12}
\tilde L\m=\tilde S(x)\big(1+ia\tilde A_\mu(x)-a\tilde C_\mu(x)\big).
\ee
Here $\tilde S$ and $\tilde C_\mu$ play the role of ``matter fields'' with homogeneous gauge transformation properties, while $\tilde A_\mu$ are gauge fields. 
We will discuss the role of the vector bilinear $\tilde C_\mu$  in more detail below. 

In the lattice version a local gauge transformation reads
\be\label{80F7}
\delta\tilde B_\mu(x)=i[\alpha(x),\tilde B_\mu(x)]-\tilde L\m\partial_\mu\alpha(x).
\ee
This guarantees that $\tilde L\tilde L^\dagger$ transforms homogeneously,
\be\label{80F9}
\delta(\tilde L\tilde L^\dagger)=i[\alpha,\tilde L\tilde L^\dagger].
\ee
The definition of $\tilde A_\mu$ and its transformation properties have to be adapted correspondingly. 
In the view of the continuum properties we propose in the lattice formulation a decomposition similar to eq. \eqref{19} 
\be\label{80F13}
\tilde L\m=\tilde S(x)\big(1-a\tilde C_\mu(x)\big)\exp \big\{ia\tilde A_\mu(x)\big\}.
\ee
The local gauge transformation of $\tilde A_\mu$ is given implicitly by 
\ba\label{80F14}
\delta \exp\big (ia\tilde A_\mu(x)\big)&=&i\big[\alpha(x),\exp\big(ia\tilde A_\mu(x)\big)\big]\\
&&-ia\exp \big (ia\tilde A_\mu(x)\big)\partial_\mu\alpha(x),\nn
\ea
which becomes eq. \eqref{80F6} in lowest order in $a$. Inserting also the transformations \eqref{80F4} and \eqref{80F11} for $\tilde S$ and $\tilde C_\mu$ one establishes
\be\label{80F15}
\delta\tilde L\m=i\big[\alpha(x),\tilde L\m\big]-ia\tilde L\m\partial_\mu\alpha(x),
\ee
and therefore eq. \eqref{80F9}. 

One can find an alternative expression of $\tilde L$ as
\ba\label{IA1}
\tilde L\m&=&\chi(x+e_\mu)\chi^\dagger(x+e_\mu)\nn\\
&&-\big[\chi(x+e_\mu)
-\chi(x)\big]\chi^\dagger(x+e_\mu)\nn\\
&=&\tilde S(x+e_\mu)+ia\tilde B^{'\dagger}_\mu(x+e_\mu),
\ea
with 
\be\label{IA2}
\tilde B'_\mu(x)=-\frac ia\chi(x)\big(\chi^\dagger(x)-\chi^\dagger(x-e_\mu)\big).
\ee
Correspondingly, we define on the lattice $\tilde C'_\mu$ similar to eq. \eqref{80F13} by
\ba\label{IA4}
\tilde L^\dagger\m&=&\tilde S(x+e_\mu)-ia\tilde B'_\mu(x+e_\mu)\nn\\
&=&\tilde S(x+e_\mu)\big(1+a\tilde C'_\mu(x+e_\mu)\big)\nn\\
&&\times\exp \{-ia\tilde A_\mu(x+e_\mu)\}\nn\\
&=&\tilde L(x+e_\mu;-\mu).
\ea
A $\pi$-rotation results in $\tilde L\m\to\tilde L(x+e_\mu;-\mu)$ and therefore in $\tilde A_\mu\leftrightarrow -\tilde A_\mu$, $\tilde C_\mu\leftrightarrow-\tilde C'_\mu$, while $\tilde S$ is invariant. 
The alternative representation of the link bilinear will be useful when we discuss the continuum fields of our model. 

\medskip\noindent
{\bf 2. Lattice rotations and reflections}

Lattice rotations by $\pi/2$ and coordinate reflections act on the scalars $\chi$ in the usual way, $\chi(x)\to\chi(x')$, with $x'$ the image of $x$ under rotations or reflections. 
This extends to the scalars $M$ or $\tilde S$ which are constructed from scalars at the same lattice point. 

Consider next the transformation of link variables under $\pi/2$-rotations in the $\mu-\nu$-plane around the center of a plaquette at $y=x+(e_\mu+e_\nu)/2$. 
The link bilinears of this plaquette transform as
\ba\label{TT1}
\tilde L\m&\to&\tilde L(x+e_\mu;\nu)\to\tilde L^\dagger(x+e_\nu;\mu)\nn\\
&\to&\tilde L^\dagger(x;\nu)\to\tilde L\m. 
\ea
With the ansatz \eqref{80F13}, \eqref{IA4} this implies
\ba\label{TT2}
\tilde S(x)&\to&\tilde S(x+e_\mu)\to\tilde S(x+e_\mu+e_\nu)\nn\\
&\to&\tilde S(x+e_\nu)\to\tilde S(x),\nn\\
\tilde A_\mu(x)&\to& \tilde A_\nu(x+e_\mu)\to-\tilde A_\mu(x+e_\mu+e_\nu)\nn\\
&\to& -\tilde A_\nu(x+e_\nu)\to\tilde A_\mu(x),\nn\\
\tilde C_\mu(x)&\to&\tilde C_\nu(x+e_\mu)\to-\tilde C'_\mu(x+e_\mu+e_\nu)\nn\\
&\to&-\tilde C'_\nu(x+e_\nu)\to\tilde C_\mu(x).
\ea
Here we have employed eq. \eqref{80F13} for $\tilde L\m$ and $\tilde L(x+e_\mu;\nu)$, and the alternative representation \eqref{IA4} for $\tilde L^\dagger(x+e_\nu;\mu)$ and $\tilde L^\dagger(x;\nu)$. 
We recognize the standard scalar or vector transformation properties for $\tilde S$ and $\tilde A_\mu$, respectively. Furthermore, the combinations 
\ba\label{TT3}
\bar C_\mu(y)&=&\frac12\big(\tilde C_\mu(x)+\tilde C'_\mu(x+e_\mu+e_\nu)\big),\nn\\
\bar C_\nu(y)&=&\frac12\big(\tilde C_\nu(x+e_\mu)+\tilde C'_\nu(x+e_\nu)\big),
\ea
transform as the components of a vector. In the continuum limit we will therefore associate both $\tilde A_\mu$ and $\bar C_\mu$ with a vector. 

For a reflection of the coordinate $x^\mu$ at a plane going through the center of the cell one has $x\leftrightarrow x+e_\mu$, $x+e_\nu\leftrightarrow x+e_\mu+e_\nu$, and correspondingly
\ba\label{116A}
\tilde L\m&\leftrightarrow& \tilde L^\dagger \m,\nn\\
\tilde L(x;\nu)&\leftrightarrow&\tilde L(x+e_\mu;\nu).
\ea
For $\tilde A_\mu$ this entails the standard transformation properties of a parity odd vector field 
\be\label{116B}
\tilde A_\mu(x)\leftrightarrow -\tilde A_\mu(x+e_\mu)~,~\tilde A_\nu(x)\leftrightarrow\tilde A_\nu(x+e_\mu).
\ee
For the bilinears $\tilde C_\mu,\tilde C'_\mu$ one finds
\ba\label{116C}
\tilde C_\mu(x)&\leftrightarrow&-\tilde C'_\mu(x+e_\mu)~,~\tilde C_\nu(x)\leftrightarrow \tilde C_\nu(x+e_\mu),\nn\\
\tilde C'_\nu(x)&\leftrightarrow&\tilde C'_\nu(x+e_\mu),
\ea
and therefore 
\ba\label{116D}
\bar C_\mu(\tilde y)&\to&-\frac12\big(\tilde C_\mu(x+e_\nu)+\tilde C'_\mu(x+e_\mu)\big),\nn\\
\bar C_\nu(\tilde y)&\to&\frac12\big(\tilde C_\nu(x)+\tilde C'_\nu(x+e_\mu+e_\nu)\big).
\ea
Standard (parity odd) vector properties of the field $\bar C_\mu(\tilde y)$ obtain only 
if we assume the leading order relations $\tilde C_\mu(x+e_\nu)\approx \tilde C_\mu(x)$, $\tilde C'_\mu(x+e_\mu)\approx \tilde C'_\mu(x+e_\mu+e_\nu)$. 

\medskip\noindent
{\bf 3. Continuum limit of scalar lattice gauge theory}

The continuum limit of scalar lattice gauge theory in terms of the gauge invariant ``meson bilinears'' $M(x)$ is discussed in appendix A. 
We supplement this here by a continuum formulation which takes the collective gauge fields $\tilde A_\mu$ explicitly into account. 
This version exhibits more clearly the features that are relevant for the true continuum limit of our model. 
We define the continuum limit by assuming that the lattice fields $\chi(x)$ are sufficiently smooth such that lattice derivatives can be replaced by partial derivatives of corresponding continuum fields. 
(If we want to include configurations with strong variation, say of an antiferromagnetic type, the continuum limit would have to be extended by the inclusion of additional fields.)

In the continuum limit $\tilde B'_\mu(x)$ coincides with $\tilde B_\mu(x)$, 
\be\label{116E}
\tilde B_\mu(x)=\tilde B'_\mu(x)=-i\chi(x)\partial_\mu\chi^\dagger(x).
\ee
Comparing eqs. \eqref{IA1} and \eqref{80F1} in leading order in $a$ gives rise to the identity
\ba\label{IA3}
\partial_\mu\tilde S&=&-i(\tilde B^\dagger_\mu-\tilde B_\mu)\nn\\
&=&-i[\tilde A_\mu,\tilde S]-\{\tilde C_\mu,\tilde S\}.
\ea
This relates the covariant derivative of $\tilde S$ to the anticommutator $\{\tilde C^\dagger_\mu,\tilde S\}$,
\ba\label{117A}
\frac12\{\tilde C_\mu,\tilde S\}&=&-\frac12 D_\mu\tilde S,\nn\\
D_\mu\tilde S&=&\partial_\mu\tilde S+i[\tilde A_\mu,\tilde S].
\ea
As a consequence, we may express $\tilde C_\mu$ in terms of $\tilde S$ and its covariant derivative. 

Let us next discuss the continuum limit of $\tilde L\m$. In analogy to eq. \eqref{19} we may write
\be\label{117B}
\tilde L\m=\bar S\m\tilde U\m.
\ee
Using the decomposition \eqref{80F13} we identify 
\ba\label{117C}
\tilde U\m&=&\exp\big (ia\tilde A_\mu(x)\big),\nn\\
\bar S\m&=&\tilde S(x)\big (1-a\tilde C_\mu(x)\big).
\ea
We next employ eq. \eqref{80F10} 
\ba\label{117D}
\tilde S\tilde C_\mu=-i(\tilde B_\mu-\tilde S\tilde A_\mu)=-\chi(D_\mu\chi)^\dagger,
\ea
with covariant derivative
\be\label{117E}
D_\mu\chi=\partial_\mu\chi+i\tilde A_\mu\chi.
\ee
This combines to
\ba\label{117F}
\bar S\m&=&\tilde S(x)+a\chi(x)D_\mu\chi^\dagger(x)\nn\\
&=&\chi(x)\chi^\dagger(x)+a\chi(x)D_\mu\chi^\dagger(x).
\ea

Potential terms for links are functions of $\tilde L\m\tilde L^\dagger\m=\bar S\m\bar S^\dagger\m$, with 
\ba\label{117G}
&&\tilde L\m\tilde L^\dagger\m=\tilde S^2(x)+\chi(x)\big[a\partial_\mu\big (\chi^\dagger(x)\chi(x)\big)\nn\\
&&\hspace{1.5cm}+a^2D_\mu\chi^\dagger(x)D_\mu\chi(x)\big]\chi^\dagger(x).
\ea
For example, one has 
\ba\label{117H}
&&\tr\big\{\tilde L\m\tilde L^\dagger\m\big\}=\tr \tilde S^2(x)+
\frac a2\partial_\mu\tr M^2(x)\nn\\
&&\qquad +a^2\tr \big\{\chi^\dagger(x)\chi(x)D_\mu\chi^\dagger(x)D_\mu\chi(x)\big\}.
\ea
In the sum over all links, $\sum_{\rm links} \tr\{\tilde L\tilde L^\dagger\}$, the term $\sim\partial_\mu\tr M^2$ drops out (or amounts at most to a boundary term), 
such that the leading contributions are rotation invariant. We observe the relations
\be\label{117I}
\tr \tilde S^K(x)=\tr M^K(x),
\ee
which identifies the singlet $\tr \tilde S=\tr M$ and links the adjoint representation in $\tilde S$ to gauge invariant meson bilinears.

\section*{APPENDIX D: LATTICE ACTION FOR SCALARS AND LINK VARIABLES}
\renewcommand{\theequation}{E.\arabic{equation}}
\setcounter{equation}{0}

In this appendix we discuss in more detail the generalized Hubbard-Stratonovich transformation that links the action of scalar lattice gauge theory to the one of the equivalent link-scalar model. 
We also extend our discussion to actions that generalize the simple action \eqref{30C}.
For this purpose we generalize $W_L(L)$ in eq. \eqref{18} to
\be\label{EA}
W_L(L) = -\mu^2\rho + \frac{\lambda_1}{2}\rho^2+\frac{\lambda_2}{2}\tau.
\ee
The link action in sect. \ref{Gauge bosons as collective excitations} corresponds to $\lambda_1 = \lambda_2/2 = \lambda$.
We also add terms to the scalar action, generalizing $S_\rchi$ beyond eq. \eqref{30C}.

\medskip\noindent
{\bf 1. Action for the link-scalar model}

Insertion of eq. \eqref{A2} into the scalar functional integral \eqref{30} yields a lattice model for link variables and scalar variables. The microscopic action for this ``link-scalar model'' consists of four pieces,
\ba\label{A2A}
S[\chi,L]=S_\chi+S_L+\Delta S_\chi+S_{{\rm int}}:
\ea
(i) The scalar action $S_\chi[\chi]$ is the original action of scalar lattice gauge theory in eq. \eqref{30}. We take it in the form
\ba\label{A3}
S_\chi[\chi]&=&S_V[\chi]+S_{{\rm kin}}[\chi]+\sum_{{\rm links}}W_\chi[\tilde L]\nn\\
&&+\sum_{\rm plaquettes}\big({\cal A}_p[\chi]-{\cal L}_p[\tilde L]\big).
\ea
The last term corresponds to $\sum{\cal S}_p$ in eqs. \eqref{M1}, \eqref{M3} with ${\cal L}_p$ and ${\cal A}_p$ given by eqs. \eqref{M2} and  \eqref{M4}. The potential term $S_V$ (cf. eq. \eqref{2}) 
\be\label{61A}
S_V=\sum_xV_\chi\big(\chi(x)\big),
\ee
and kinetic term $S_K$ (similar to eq. \eqref{4}), as well $W_\chi[\tilde L]$, generalize the scalar link action ${\cal S}_l$ in \eqref{30D}. They will be determined later. 
In other words, the sum $S_V+S_{\rm kin}+\sum W_\chi$ will correspond to $\bar S_l$ in eq. \eqref{30C} or generalizations thereof. 
(ii) The link action $S_L[L]$ is the part of $\bar S_L$ in eq. \eqref{A2} that only involves $L$, while 
(iii) the shift in the scalar action $\Delta S_\chi[\chi]$ is the corresponding part from eq. \eqref{A2} that only involves $\chi$. 
(iv) Finally, there is a coupling between the links and the scalars, given by eq. \eqref{A4}.

The particular form of the transformation \eqref{A2} has been chosen in order to eliminate the pieces $\sim -{\cal L}_p$ and $-{\cal L}_l$, 
thereby rendering the residual scalar part of the action $S_\chi+\Delta S_\chi$ much simpler. 
First, the plaquette term $S_p [\tilde L]$ in $\Delta S_\chi$ cancels the plaquette term $-S_p[\tilde L]$ in $S_\chi$. 
This eliminates the most cumbersome derivative interaction in scalar lattice gauge theory in favor of a plaquette action for links. 
Furthermore, by the use of identities of the type \eqref{8} or \eqref{16B} we can always write $S_\chi$ in the form \eqref{A3} with $W_\chi[\tilde L]=-W_L[-\tilde L]$. 
Thus $W_\chi[\tilde L]$ is canceled by the contribution $W_L[-\tilde L]$ from $\Delta S_\chi$. 
The microscopic action for scalars and links becomes then
\be\label{A6}
S[\chi,L]=S_V[\chi]+S_{{\rm der}}[\chi]+S_L[L]+S_{{\rm int}}[\chi,L],
\ee
where the derivative terms for $\chi$ include interactions
\be\label{56A}
S_{\rm der}[\chi]=S_{\rm kin}[\chi]+\sum_{\rm plaquettes}{\cal A}_p.
\ee
This action describes linear lattice gauge theory coupled to scalars.  

The remaining derivative part $S_{\rm der}[\chi]$ is much simpler than the derivative part in the original action $S_\chi$ of scalar lattice gravity. 
As we have discussed in sect. \ref{Invariant action and} the invariant ${\cal A}_p$ suppresses the contributions of highly inhomogeneous scalar bilinears $M(x)$ to the functional integral. 
The price to pay is the presence of explicit link variables for the gauge field and a relatively complex interaction term. 
A simpler picture is only realized if the role of the interaction term is subleading, and we have discussed conditions for this in section \ref{Phase diagram of scalar}.

As an example, the contribution $\sim \mu^2$ to the interaction term reads in the matrix notation
\be\label{FB}
S^{(\mu^2)}_{\rm int}=\mu^2\sum_{\rm links}{\rm tr}\big\{\chi^\dagger(x+e_\mu)
L^\dagger(x;\mu)\chi(x)\big\}+c.c..
\ee
If we define a covariant derivative $D_\mu$ by
\be\label{FC}
a{D}_\mu\chi(x)=L(x;\mu)\chi(x+e_\mu)-\chi(x)
\ee
we can write invariants of the type 
\ba\label{FD}
&&{\rm tr} \big \{\big(D_\mu\chi(x)\big)^\dagger D_\mu \chi(x)\big\}=\nn\\
&&\quad \tr\big\{\chi^\dagger(x+e_\mu)L^\dagger(x;\mu)L \m\chi(x+e_\mu)\\
&&\quad +\chi^\dagger(x)\chi(x)-\big(\chi^\dagger(x+e_\mu)L^\dagger\m\chi(x)+c.c.\big )\big\}.\nn
\ea
We can associate the interaction \eqref{FB} with the interaction between two scalars and a link that appears in the squared covariant derivative. 

\medskip\noindent
{\bf 2. Action for scalar lattice gauge theory}

We next discuss more specifically the action \eqref{A3} for scalar lattice gauge theory that is equivalent to the link-scalar action \eqref{A6}. 
We will show that for a suitable choice of $S_V$ and $S_{\rm kin}$ and for the choice of couplings $\lambda_1=\lambda_2/2$ 
the action $S_\chi$ \eqref{A3} equals the scalar lattice action \eqref{30C} of sect. \ref{Invariant action and}, 
thereby establishing the equivalence of the ``link-scalar model'' \eqref{A6} with the scalar lattice gauge theory of sect. \ref{Invariant action and}. 
For more general choices we obtain extended models with similar properties.

In order to make this connection we have to express $W_\chi[\tilde L]$ in terms of $M$ and to specify $S_V$ and $S_{\rm kin}$. For the choice \eqref{EA} one has
\ba\label{A9}
W_\chi(\tilde L)&=&-W_L(-\tilde L)=-W_L(\tilde L)\nn\\
&& =\mu^2\tilde \rho-\frac{\lambda_1}{2}\tilde \rho^2-\frac{\lambda_2}{2}\tilde \tau_2.
\ea
Here $\tilde \rho$ and $\tilde \tau_2$ obey eq. \eqref{18} with the replacement $L\to\tilde L$. In terms of the scalar matrix $M$ they read 
\ba\label{A10}
\tilde \rho\m&=&{\rm tr}\big\{M(x)M(x+e_\mu)\big\},\nn\\
\tilde \tau_2\m&=&\frac N2{\rm tr}\Big\{\big[M(x)M(x+e_\mu)\nn\\
&&-\frac1N{\rm tr}\big\{M(x)M(x+e_\mu\big\}\big]^2\Big\}.\nn
\ea

We write
\be\label{56B}
\sum_{\rm links}W_\chi[\tilde L]=S_{W,{\rm der}}+S_{W,l}-\sum_x V_W(x)
\ee
with 
\ba\label{A13}
&&S_{W,{\rm der}}=\frac{a^2}{2}\sum_x\sum_\mu\Big[\left(\frac{\lambda_1}{4}-\frac{\lambda_2}{8}\right)\\
&&\times \Big\{\Big[{\rm tr}\big\{\big[M(x)+M(x+e_\mu)\big]\partial_\mu M(x)\big\}\Big]^2\nn\\
&&+{\rm tr}\big\{\big[M(x)+M(x+e_\mu)\big]^2\big\}{\rm tr}\big\{\partial_\mu M(x)\partial_\mu M(x)\big\}\Big\}\nn\\
&&+\frac{\lambda_2N}{4}{\rm tr}\big\{\big[M(x)+M(x+e_\mu)\big]^2\partial_\mu M(x)\partial_\mu M(x)\big\}\Big],\nn
\ea
and 
\be\label{56C}
S_{W,l}=\sum_{\rm links}\frac{\mu^2}{2}{\rm tr}\big\{[M(x+e_\mu)+M(x)]^2\big\}.
\ee
The corresponding potential term reads
\ba\label{A12}
V_W(\chi)&=&d\mu^2{\rm tr} M^2(x)+\frac{d}{2}(\lambda_1-\frac12\lambda_2)\big({\rm tr}M^2(x)\big)^2\nn\\
&&+\frac{dN\lambda_2}{4}{\rm tr} M^4(x).
\ea

Defining for $S_V$ in eq. \eqref{A3}, \eqref{61A}
\be\label{A11}
V_\chi(\chi)=\bar V(\chi)+V_W(\chi),
\ee
we arrive at the action \eqref{A3} for scalar lattice gauge theory 
\be\label{A15}
S_\chi[\chi]=\sum_x\bar V\big(\chi(x)\big)+S_{W,l}+\bar S_{{\rm der}},
\ee
with derivative term composed from three pieces
\be\label{A16}
\bar S_{\rm der}=S_{\rm kin}[\chi]+S_{W,{\rm der}}+\bar S_p.
\ee
For a given choice of $\bar V(\chi)$ and $S_{\rm kin}$ this fixes the action $S_\chi[\chi]$ of lattice scalar gravity completely. 
Eqs. \eqref{A15}, \eqref{A16} are our final result for the choice of $S_\chi$. 
We observe that $S_{W,l}$ can be written as a sum of potential and derivative terms for $M$. 

At this point we may compare with the action \eqref{30C} in sect. \ref{Invariant action and}. 
The term $\bar S_p$ is common. In $S_{W,{\rm der}}$ the last term equals the first term in eq. \eqref{30D} if we relate $\bar\lambda$ and $\lambda_2=2\lambda$ by eq. \eqref{XAX}.
Furthermore $S_{W,{\rm der}}$ contains an additional derivative term $\sim (\lambda_1-\lambda_2/2)$ that we have omitted in sect. \ref{Invariant action and} (e.g. setting $\lambda_1=\lambda_2/2$). 
The contribution $S_{W,l}$ equals the second term in eq. \eqref{30D} provided we use the relation \eqref{XAY}.

In summary, the action \eqref{A3} of scalar lattice gauge theory equals the action \eqref{30C}, plus additional pieces
\be\label{A74}
S'=\sum_x\bar V\big(\chi(x)\big)+S_{\rm kin}+\Delta S_{\rm der},
\ee
with $\Delta S_{\rm der}$ the term in $S_{W,{\rm der}}$ proportional $(\lambda_1-\lambda_2/2)$. 
For a simplified discussion we may take $S'=0$, such that eqs. \eqref{30C} and \eqref{A15} coincide. 
Scalar lattice gauge theory  with action $S_\chi[\chi]$ given by eq. \eqref{30C} is then completely equivalent to the link-scalar model with functional integral over links and scalars and action $S[\chi,L]$ given by eq. \eqref{A6}. 
For the particular lattice gauge theory of sect. \ref{Invariant action and} the equivalent link-scalar model has the action \eqref{84A}.

\section*{APPENDIX E: CONNECTION BETWEEN LINK VARIABLES AND LINK BILINEARS}
\renewcommand{\theequation}{F.\arabic{equation}}
\setcounter{equation}{0}

The functional integrals of scalar lattice gauge theory with action \eqref{30C} and for the link-scalar model with action \eqref{84A} are equivalent. 
This does not change if we add sources as in eq. \eqref{30}. Thus the effective action \eqref{34} is the same, and we just deal with two different ways to compute it. 
In the link-scalar model we can, in addition, compute the expectation value $\kl L\m\kr$ or correlation functions for the link variables. 
The question arises how these quantities are related to the expectation value of the link bilinear $\kl \tilde L\m\kr$ and corresponding correlation functions. 

\medskip\noindent
{\bf 1. Source-dependent Hubbard-Stratonovich trans-

\hspace{0.2cm}formation}

At the present stage the collective source $K$ multiplies only the bilinear $\tilde L$ according to eq. \eqref{30}. 
A coupling of $K$ to the link variables $L$ can be implemented by a modification of the Hubbard-Stratonovich type transformation. 
We change from the choice $\bar S_L=S_L[L-\tilde L]$ to $\bar S_L= S_L[L-\tilde L-\hat K]$, with $\hat K$ a linear combination of the different representations in $K$,
\ba\label{64A}
K&=&k_R+ik_I+K_A+iK_B,\nn\\
\hat K&=&c_Rk_R+ic_Ik_I+c_AK_A+ic_BK_B.
\ea
Here $c_R,c_I, c_A, c_B$ are real constants, $k_R$ and $k_I$ are proportional to the unit matrix, and $\tr K_A=\tr K_B=0,K^\dagger_A=K_A,K^\dagger_B=K_B$. The factor $Z_L$ in eq. \eqref{A2} is the same, such that the model is not changed for an arbitrary choice of coefficients $c_j$. The functionals $W[j,K]$ and $\Gamma[\varphi,\bar L]$ are the same for scalar lattice gauge theory and the link-scalar model obtained from the $K$-dependent modification of the transformation \eqref{A2}. However, the association of $K$-derivatives of $W$ with expectation values and correlation functions depends on the choice of $c_j$. 

This can seen by expanding $S_L[L-\tilde L-\hat K]$ around its minimum  at $L-\tilde L-\hat K=l_0$. With the decomposition
\ba\label{64B}
L&=&l+is+L_A+iL_B,\nn\\
l&=&\frac{1}{2N}\tr \{L+L^\dagger\}~,~s=-\frac{i}{2N}\tr\{L-L^\dagger\},\nn\\
L_A&=&\frac12(L+L^\dagger)-l~,~L_B=-\frac i2(L-L^\dagger)-s,\nn\\
\tr L_A&=& \tr L_B=0~,~L^\dagger_A=L_A~,~L^\dagger_B=L_B,
\ea
and similar for $\tilde L$, one finds (omitting an irrelevant constant)
\ba\label{S1}
S_L&=&\sum_{\rm links}2N^2\lambda_1 l^2_0(l-\tilde l-l_0-c_Rk_R)^2\\
&&+N\lambda_2l^2_0\tr\{(L_A-\tilde L_A-c_AK_A)^2\} +S^{(2)}_{L{\rm der}}+\dots\nn
\ea
where $S^{(2)}_{L{\rm der}}$ contains derivative terms from an expansion of $S_p$ in quadratic order in $(L-\tilde L-\hat K-l_0)$.
(We take in this appendix the more general form \eqref{EA} for $W_L$. The model in the main text corresponds to $\lambda_1=\lambda_2/2=\lambda$.)

Neglecting the derivative terms and higher orders we find a contribution of $S_L$ linear in $k_R$ and $K_A$
\ba\label{S2}
S^{(1)}_{L,K}&=&-4N^2\lambda_1 l^2_0c_Rk_R(l-\tilde l-l_0)\nn\\
&&-2N\lambda_2l^2_0c_A\tr\{K_A(L_A-\tilde L_A)\}.
\ea

This yields for every link $\m$
\ba\label{S3}
\frac{\partial W}{\partial k_R}&=&2N\kl \tilde l+2N\lambda_1 l^2_0 c_R(l-\tilde l-l_0)\kr,\nn\\
\frac{\partial W}{\partial K_A}&=&2\kl \tilde A+N\lambda_2 l^2_0c_A(L_A-\tilde L_A)\kr,
\ea
to be compared with the equivalent eq. \eqref{31} which corresponds to $c_R=c_A=0$. Here we have neglected contributions for $K\neq 0$ which will modify higher order correlation functions. For $c_j\neq 0$ the expectation value $\kl \tilde L\kr_\chi$ in the scalar lattice gauge theory does not equal the value $\kl \tilde L\kr_{\chi L}$ in the equivalent link-scalar model. This is due to the additional expectation value $\kl L\kr_{\chi L}$ of the link variable. In particular, for the choice $c_R=(2N\lambda_1 l^2_0)^{-1}$ the collective source $k_R$ decouples from the link bilinear $\tilde l$ and one obtains the leading order relation 
\be\label{S4}
\kl \tilde l\kr_\chi=\kl l\kr_{\chi L}-l_0.
\ee
Similarly, for $c_A=(N\lambda_2l^2_0)^{-1}$ the source $K_A$ decouples from $\tilde A$ and one has
\be\label{67A}
\kl\tilde L_A\kr_\chi=\kl L_A\kr_{\chi L}.
\ee

Eq. \eqref{S4} demonstrates that the expectation values of link bilinears in the link-scalar model $\kl \tilde L\kr_{\chi L}$ need not to be the same as the ones for the original scalar model $\kl \tilde L\kr_\chi$. In general, they will depend on the parameters $c_j$ used in the Hubbard-Stratonovich type transformation. Eqs. \eqref{S4}, \eqref{67A} also demonstrate that it is possible to express $\kl \tilde L\kr_\chi$ in terms of expectation values of link variables in the link-scalar model. We learn from eq. \eqref{S4} that the expectation value of the link variable in the link-scalar model $\kl L \kr_{\chi L}$ can differ from zero even if the expectation value of the link bilinear $\kl \tilde L\kr_\chi$ vanishes in the original scalar model. 

The relations \eqref{S4}, \eqref{67A} get modified by higher orders in the expansion of $S_L$ in powers of $(L-\tilde L-l_0-\hat K)$. The terms linear in $K$ add on the r.h.s. expectation values $\sim\kl (L-\tilde L-l_0)^n\kr,n=2,3$. Higher powers in $K$ in the expansion of $S_L[L-\tilde L-\hat K]$ will influence the precise relation between correlation functions in scalar lattice gauge theory and the equivalent link-scalar model. The situation is similar for the sources $k_I$ and $K_B$ for which the additional contributions arise from $S_p$. One can choose coefficient $c_j$ such that for vanishing sources $K=0$ the relations \eqref{S4}, \eqref{67A} are obeyed, together with similar relations as $\kl \tilde L_B\kr_\chi=\kl L_B\kr_{\chi L}$. In Fourier space the coefficients $c_j$ can be taken as functions of the squared momentum.

The upshot of these considerations shows that the correlation functions for the link variables $L$ in the link-scalar model have similar properties as corresponding correlation functions for $\tilde L$ in scalar lattice gauge theories. The detailed relation is rather involved, however, and reflects operator mixing for observables with the same transformation properties. Despite these complications the expectation values of Wilson loops for $L$ can be used for confinement criteria similar to standard lattice gauge theories. 

\medskip\noindent
{\bf 2. Equivalent link-scalar models}

We emphasize that the choice $\bar S_L=S_L$ is only one particular possibility to introduce an integration over link variables. Many other choices are possible and may be convenient in order to obtain a simpler form for the interaction $S_{{\rm int}}$. As an example, consider 
\be\label{A7}
\bar S_L=S_{LK}=\sum_{{\rm links}}{\rm tr}(L^\dagger -\tilde L^\dagger-K^\dagger)(L-\tilde L-K).
\ee
This results in 
\ba\label{A8}
S_L&=&\sum_{{\rm links}}{\rm tr}\{L^\dagger L\}~,~S_{{\rm int}}=-\sum_{{\rm links}} {\rm tr}\{L^\dagger\tilde L+\tilde L^\dagger L\},\nn\\
\Delta S_\chi&=&\sum_{{\rm links}} {\rm tr}\{\tilde L^\dagger\tilde L\}.
\ea
Furthermore, the term tr$\big\{(\tilde L^\dagger-L^\dagger)K+K^\dagger(\tilde L-L)\big\}$ replaces in eq. \eqref{30} the source term for the bilinear $K^\dagger\tilde L$ by a standard linear source term for the links $K^\dagger L$, resulting in $\kl L\kr_{\chi L}=\kl\tilde L\kr_\chi$. (The additional term in $W\sim$tr$K^\dagger K$ has to be taken into account if correlation functions are computed from derivatives of $W$.)

With the choice \eqref{A7} the interaction becomes a simple cubic interaction involving one link and two scalar fields $\sim$tr$L^\dagger\chi\chi^\dagger$. The microscopic link action contains only a quadratic link potential.  In particular, there is no plaquette term ${\cal L}_p$ for the links, while the cumbersome derivative interaction $-{\cal L}_p(M)$ is still present in the scalar sector. Nevertheless, the two choices $\bar S_L=S_L$ or $\bar S_L=S_{LK}$ are completely equivalent. For the choice $\bar S_L=S_{LK}$ an effective plaquette term for the link variables $L$ is expected to be induced by quantum fluctuations.

\bibliography{scalar_lattice_gauge_theory}

\end{document}